\pdfoutput=1
\documentclass[usenatbib,usegraphicx]{mn2e}
\usepackage{amssymb}
\usepackage{graphicx}
\usepackage{epsfig}
\usepackage{color}
\usepackage{longtable}

\newcommand{\OVIdblt}{{O}\kern 0.1em{\sc vi}~$\lambda\lambda 1032, 1038$}
\newcommand{\CII}{\hbox{{C}\kern 0.1em{\sc ii}}}
\newcommand{\CIII}{\hbox{{C}\kern 0.1em{\sc iii}}}
\newcommand{\CIV}{\hbox{{C}\kern 0.1em{\sc iv}}}
\newcommand{\HI}{\hbox{{H}\kern 0.1em{\sc i}}}
\newcommand{\Lya}{\hbox{{Ly}\kern 0.1em$\alpha$}}
\newcommand{\Lyb}{\hbox{{Ly}\kern 0.1em$\beta$}}
\newcommand{\Lyg}{\hbox{{Ly}\kern 0.1em$\gamma$}}
\newcommand{\Lyd}{\hbox{{Ly}\kern 0.1em$\delta$}}
\newcommand{\Lye}{\hbox{{Ly}\kern 0.1em$\epsilon$}}
\newcommand{\Lyz}{\hbox{{Ly}\kern 0.1em$\zeta$}}
\newcommand{\Lyeta}{\hbox{{Ly}\kern 0.1em$\eta$}}
\newcommand{\MgII}{\hbox{{Mg}\kern 0.1em{\sc ii}}}
\newcommand{\OVI}{\hbox{{O}\kern 0.1em{\sc vi}}}
\newcommand{\OVII}{\hbox{{O}\kern 0.1em{\sc vii}}}
\newcommand{\OVIII}{\hbox{{O}\kern 0.1em{\sc viii}}}
\newcommand{\NV}{\hbox{{N}\kern 0.1em{\sc v}}}
\newcommand{\SiII}{\hbox{{Si}\kern 0.1em{\sc ii}}}
\newcommand{\SiIII}{\hbox{{Si}\kern 0.1em{\sc iii}}}
\newcommand{\SiIV}{\hbox{{Si}\kern 0.1em{\sc iv}}}
\newcommand{\FeII}{\hbox{{Fe}\kern 0.1em{\sc ii}}}

\newcommand{\NeIX}{\hbox{{Ne}\kern 0.1em{\sc ix}}}
\newcommand{\OI}{\hbox{{O}\kern 0.1em{\sc i}}}

\def\apj{ApJ}

\def\mnras{MNRAS}

\title[Cool Clouds Acceleration and Destruction]
{Entrainment in Trouble: Cool Cloud Acceleration and Destruction in Hot Supernova-Driven Galactic Winds}

\author[Zhang et al.]
{Dong Zhang$^{1,2,3}$\thanks{E-mail:dz7g@virginia.edu}, Todd A.~Thompson$^{2,3}$,
Eliot Quataert$^{4}$ and Norman Murray$^{5}\thanks{Canada Research Chair in Astrophysics}$\\
$^1$Department of Astronomy, University of Virginia, 530 McCormick Road, Charlottesville, VA 22904, USA\\
$^2$Department of Astronomy, The Ohio State University, 140 West 18th Avenue, Columbus, OH 43210, USA\\
$^3$Center for Cosmology \& Astro-Particle Physics, The Ohio State University, Columbus, Ohio 43210, USA\\
$^4$Astronomy Department \& Theoretical Astrophysics Center, 501 Campbell Hall, University of California, Berkeley, CA 94720, USA\\
$^5$Canadian Institute for Theoretical Astrophysics, 60 St. George Street, University of Toronto, Toronto, ON M5S 3H8, Canada}

\begin{document}

\maketitle

\begin{abstract}
Efficient thermalization of overlapping supernovae within star-forming galaxies may produce a supernova-heated fluid that drives galactic winds. For fiducial assumptions about the timescale for cloud shredding from high-resolution simulations (which neglect magnetic fields) we show that cool clouds with temperature from $T_{c}\sim 10^{2}-10^{4}$ K seen in emission and absorption in galactic winds cannot be accelerated to observed velocities by the ram pressure of a hot wind. Taking into account both the radial structure of the hot flow and gravity, we show that this conclusion holds over a wide range of galaxy, cloud, and hot wind properties. This finding calls into question the prevailing picture whereby the cool atomic gas seen in galactic winds is entrained and accelerated by the hot flow. Given these difficulties with ram pressure acceleration, we discuss alternative models for the origin of high velocity cool gas outflows. Another possibility is that magnetic fields in cool clouds are sufficiently important that they prolong the cloud's life. For $T_{c}=10^{3}$\,K and $10^{4}$\,K clouds, we show that if conductive evaporation can be neglected, the cloud shredding timescale must be $\sim15$ and 5 times longer, respectively, than the values from hydrodynamical simulations in order for cool cloud velocities to reach those seen in observations.

\end{abstract}


\section{Introduction}

Galactic winds are ubiquitous and important in rapidly star-forming galaxies. They are a primary source of metals in the intergalactic medium and affect the chemical evolution of galaxies (e.g., \citealt{Dekel86}; \citealt{Aguirre01}; \citealt{Finlator08}; \citealt{Peeples11}).

Several mechanisms have been proposed for launching galaxy-scale outflows.  Among them, the very hot wind created by supernova (SN) energy injection is widely used in the literature. \cite{CC85} (hereafter CC85) developed a one-dimensional model for SN-driven winds with two controlling parameters: the thermalization efficiency with which SN-injected energy is converted into thermal energy, and the mass-loading efficiency, i.e., the ratio of the hot gas mass loss rate ($\dot{M}_{\rm hot}$) to the host galaxy star formation rate (SFR): $\beta=\dot{M}_{\rm hot}/$SFR. These two parameters are difficult to determine observationally. For example, observational constraints on $\beta$ have been determined for only a few galaxies, e.g., NGC 1569 (\citealt{Martin02}) and M82 (\citealt{SH09}).

In \cite{Zhang14} we derived a general constraint on $\beta$ across a wide range of galaxies from dwarf starbursts to ultra-luminous infrared galaxies (ULIRGs) using the observed  linear relation between the X-ray luminosity ($L_{X}$) and SFR (e.g., \citealt{Mineo12,Lehmer10,Mineo14}). In contrast with the observations, the CC85 model predicts $L_{X}\propto$ SFR$^{2}$ for the hot wind fluid if $\beta$ is a constant. Thus the observed $L_{X}-$SFR relation can be used to constrain the hot wind. By combining the CC85 model with a band-dependent calculation of the X-ray emission and comparing with recent determinations of the $L_{X}-$SFR relation \citep{Mineo14} we showed that $\beta\lesssim 1$ for SFR $\gtrsim 10\;M_{\odot}$ yr$^{-1}$. Larger values of $\beta$ would overproduce X-rays.

This constraint on the hot wind outflow rate implies that the CC85 model alone cannot explain the $\beta\sim1-10$ required by integrated constraints on stellar feedback models in a cosmological context (i.e., \citealt{Oppenheimer06, Oppenheimer08}; \citealt{Finlator08}; \citealt{Bower12}; \citealt{Puchwein13}). However, galactic winds are known to be multi-phase, with clear evidence for neutral atomic and ionized gas in emission and absorption from multi-wavelength observations. For example, Na I D absorption-line surveys provide the kinematics of neutral atomic outflows in local starbursts and high-$z$ star-forming galaxies (e.g., \citealt{Heckman00}; \citealt{Rupke02, Rupke05a, Rupke05b, Rupke05c}; \citealt{Schwartz04}; \citealt{Martin05}; \citealt{Weiner09}; \citealt{Erb12}; \citealt{Kornei13}). Emission lines such as H$\alpha$, N II, O II, OIII have also been used to probe cool outflowing gas in star-forming galaxies (see \citealt{Veilleux05} and reference therein). In addition, both cool and warm molecular gas are detected in outflows in some local and high-$z$ galaxies (e.g., \citealt{Sakamoto99}; \citealt{Walter02}; \citealt{Veilleux09}; \citealt{Fischer10}; \citealt{Sturm11}; \citealt{Bolatto13}; \citealt{Cicone14}). Obvious questions are whether or not the cool clouds are the dominant gas mass reservoir in the surrounding hot wind, whether or not they are accelerated by the ram pressure of the hot wind to the velocities seen, and whether or not the clouds survive the process of acceleration to both large physical scales and large velocities in order to match the spatially-resolved morphology seen in some local systems (e.g., \citealt{Heckman90}; \citealt{Heckman00}; \citealt{Martin05}; \citealt{Veilleux05}; \citealt{Leroy15}). These same issues are directly connected to the recent finding of a potentially large cool gas reservoir on 100\,kpc scales in the halos of $z\sim0$ galaxies (e.g., \citealt{werk}).

In this paper we seek general constraints on the ram pressure acceleration (RPA) of cool clouds over a broad parameter space that includes the hot wind properties (thermalization and mass-loading efficiencies), cool cloud properties (density, column density, and temperature), and galaxy properties (star formation rate, velocity dispersion of the host gravitational potential) from dwarf starbursts to ULIRGs.  Our primary goal is to assess and quantify cloud survival and acceleration in hot winds for comparison with observations of cold, cool, and warm molecular and atomic gas from $\sim10^2-10^4$\,K.\footnote{We refer to all of these clouds as ``cool'' throughout this paper unless we wish to make distinction between clouds that would be expected to be largely molecular, neutral atomic, or ionized.}

A number of studies have discussed the interaction between cool clouds and the surrounding hot outflow in rapidly star-forming galaxies.  On the observational side, the thermal soft X-ray emission shows that the hot ionized interstellar stellar medium (ISM) has a temperature of $T_{X}\sim0.2-0.8$ keV in all kinds of starburst galaxies from dwarfs to ULIRGs \citep{Martin99,Heckman00,Huo04,Grimes05}. The hot gas would be expected to accelerate cool clouds to a maximum terminal velocity of $\sqrt{3}c_{s}\simeq450(k_{B}T_{X}/0.7 $keV$)^{1/2}$ km s$^{-1}$, similar to the average velocities of cool outflows \citep{Heckman00,Rupke02,Rupke05a,Rupke05c,Martin05,Weiner09}.  On the other hand, cool gas with very high velocities above 500\,km s$^{-1}$ is also observed in some LIRGs and ULIRGs, which prima facie cannot be explained by acceleration via ram pressure of the wind that emits in soft X-rays. Thus, the very high velocity cool gas is expected to be explained by the RPA of a much hotter wind fluid associated with the diffuse hard X-ray emission. Recent observations of diffuse hard X-ray emission in M82 imply the existence of gas with $T>10^{7}$\,K (e.g., \citealt{Strickland04a}; \citealt{SH09}), which in the CC85 model would be associated with a hot wind with terminal velocity of $\sim1000-2000$ km s$^{-1}$. The H$\alpha$ filaments in M82 with velocity of $V_{\rm{H}\alpha}\sim600$\,km s$^{-1}$  (\citealt{McKeith95}; \citealt{Shopbell98}) are also proposed to be produced by RPA of cool clouds within the hot wind (e.g., \citealt{Cooper08, Cooper09}).

On the theoretical side, numerical simulations have explored both the galaxy-scale ram pressure acceleration and production of cool clouds by a hot flow (e.g., \citealt{SS00,Cooper08,Fujita09,Hopkins12}) and the survival of individual (or a set of) ram pressure accelerated clouds at high numerical resolution (e.g., \citealt{Klein94, Schiano95, Vietri97, Cooper09, Heckman00, Nakamura06, Orlando08, Jun96, Poludnenko02, Pittard05, Aluzas12}).  The galaxy-scale simulations of winds in general have coarse spatial resolution compared to what would be required to fully resolve conductive evaporation, magnetic draping, and the Rayleigh-Taylor and Kelvin-Helmholz instabilities.  Most are also tuned to one particular system (e.g., M82) and prescribe an unrealistic uniform starburst ISM as the starting condition (although, see \citealt{Cooper08,Hopkins12}).  On the other hand, although the high-resolution simulations capture much of the very small-scale physics of the clouds and their evaporation or destruction, they generically do not vary the properties of the hot wind widely or explore the evolution of the wind properties with radius as the cloud is accelerated. They also do not ask about the global effects of gravity relative to the ram pressure force, or conduct parameter studies across a wide variety of cloud properties.

In this paper, we model the dynamics of cool clouds in hot winds, varying the parameters of the problem, and tracking the dynamics of the clouds themselves, informed by the high-resolution simulations from the literature. We seek general constraints on the RPA scenario by comparing velocities, column densities, and temperatures with observations. Some quantities are given in Table \ref{definitions}. In Section \ref{Sec_Anal} we first review the hot wind solutions of CC85. We then present analytic constraints on various timescales of clouds in the hot flow, including their destruction by hydrodynamical instabilities (in particular the Kelvin-Helmholtz instability), and the acceleration timescale. We highlight the fact that acceleration timescale of the cloud is always longer than the timescale for cloud hydrodynamical instabilities, and thus the hot flow cannot accelerate cool clouds to its asymptotic velocity. We also compare the gravitational force with the ram pressure force, deriving a general Eddington-like limit as a function of cloud and host galaxy properties, which strongly constraints the initial column densities of accelerated clouds. In Section \ref{Sec_Num} we calculate cloud acceleration numerically in a spherically-symmetric model, parameterizing destruction processes and following the evolution of the cloud as it is accelerated, and as the hot wind (its density, temperature and Mach number) evolves as a function of radius. Note that a complicating factor is that the cloud destruction timescale by instabilities remains uncertain, and is a function of both the radiative properties of the cloud and its magnetization as it is crushed and accelerated by the hot flow. Recent magnetohydrodynamic simulations of isothermal clouds suggest much longer cloud lifetimes than indicated by pure hydrodynamical simulations \citep{McCourt15}. For this reason, in Section \ref{Sec_Num} we also provide additional discussion of cloud dynamics when the cloud shredding timescale is taken as a free parameter, and we derive the critical value of this timescale such that clouds are accelerated to high velocities as a guide for future simulations and comparing with observations. In Section \ref{Sec_Cases} we combine the X-ray model in \cite{Zhang14} with the RPA model for case studies of individual systems. Conclusions are presented in Section \ref{Sec_Conclusions}. We also discuss the impacts of other model parameters and other possible wind driving mechanisms.

\section{Analytic Constraints}\label{Sec_Anal}

We briefly summarize the CC85 model in this section. For more details see \cite{Zhang14}. Inside the radius of the starburst region $r\leq R$ the total energy and mass input into the hot wind are $\dot{E}_{\rm hot}$ and $\dot{M}_{\rm hot}$ and the volumetric energy and mass input rates are assumed to be constant. The flow outside the starburst region $r>R$ is assumed to be adiabatic. Under these assumptions, the Mach number $M=0$ at $r=0$, and $M=1$ at $r=R$. The two controlling dimensionless parameters of the problem, the thermalization efficiency $\alpha$  and the hot gas mass-loading efficiency $\beta$ are given by
\begin{eqnarray}
\dot{E}_{\rm hot}&=&\alpha\,\epsilon_{0}\nu_{0}\textrm{SFR},\label{parameter1}\\
\dot{M}_{\rm hot}&=&\beta\,\textrm{SFR},\label{parameter2}
\end{eqnarray}
where $\epsilon_{0}=10^{51}$\,ergs and $\nu_{0}=(100\,M_{\odot})^{-1}$ are the normalization values of the energy injected by an individual SN and the number of SNe per unit mass of star formation respectively. The temperature $T$, density $n$ and velocity $V_{\rm hot}$ of the hot wind outflow are (see \citealt{Zhang14})
\begin{eqnarray}
T(r)&=&6.1\times10^{7}\;\textrm{K}\;\mu\left(\frac{\alpha}{\beta}\right)\left[\frac{P_{*}(r_{*})}{\rho_{*}(r_{*})}\right]\label{WindT}\\
n(r)&=&14\;\textrm{cm}^{-3}\;\alpha^{-1/2}\beta^{3/2}\mu^{-1}R_{200\rm pc}^{-2}\rho_{*}(r_{*})\textrm{SFR}_{1}\label{winddensity}\\
V_{\rm hot}(r)&=&710\;\textrm{km}\;\textrm{s}^{-1}\;\alpha^{1/2}\beta^{-1/2}u_{*}(r_{*}),\label{Vhot}
\end{eqnarray}
where $R_{200\rm pc}=R/(200\,\rm{pc})$ is the wind launching radius in the host starburst, $u_{*}$, $\rho_{*}$ and $P_{*}$ are the dimensionless velocity, density and pressure as functions of the dimensionless radius $r_{*}=r/R$, and $\mu\approx0.61$ is the mean molecular weight for solar abundance.

\subsection{Initial Clouds}\label{Sec_Initial}

The dynamical timescale of the hot wind at radius $r$ is
\begin{equation}
t_{\rm dyn}\approx\frac{r}{V_{\rm hot}}\approx2.8\times10^{5}\,\textrm{yr}\;u_{*}^{-1}r_{*}\alpha^{-1/2}\beta^{1/2}R_{200 \rm pc}.\label{dyntime}
\end{equation}
The cooling timescale is
\begin{eqnarray}
t_{\rm cool}&\approx&\varepsilon_{\rm heat}/(n_{e}^{\rm hot}n_{\rm H}^{\rm hot}\Lambda_{\rm N}),\label{cooltime}
\end{eqnarray}
where $\varepsilon_{\rm heat}\approx\rho\left(\frac{1}{2}V_{\rm hot}^{2}+\frac{c_{s}^{2}}{\gamma-1}\right)=\rho_{*}\dot{E}^{1/2}\dot{M}^{1/2}/R^{2}$ is the total energy of the flow, $n_{e}^{\rm hot}$ and $n_{\rm H}^{\rm hot}$ are the electron and hydrogen density in the hot flow, and $\Lambda_{\rm N}$ is the emissivity of the flow. In \cite{Zhang14} we showed that the criterion for an adiabatic hot wind flow with $t_{\rm cool}\geq t_{\rm dyn}$ at $r=R$ implies an upper limit on $\beta$ of
\begin{equation}
\beta\leq6.6\,\alpha^{3/5}R_{200 \rm pc}^{2/5}\textrm{SFR}_{1}^{-2/5}\left(\frac{\Lambda_{\rm brems}^{\rm H}}{\Lambda_{\rm N}}\right)^{2/5},\label{constraintcooling}
\end{equation}
where SFR$_{1}$=SFR$/10\,M_{\odot}$ yr$^{-1}$, and the bremsstrahlung emission $\Lambda_{\rm brems}^{\rm H}$ is used to estimate the lower limit for the cooling rate $\Lambda_{\rm N}$, where $\Lambda_{\rm N}$ is calculated by the full SPEX package, assuming collisional ionization equilibrium and solar abundance (version 2.03.03, see \citealt{Zhang14}, also \citealt{Schure09}).  Thus, the mass loading efficiency cannot be higher than given by equation (\ref{constraintcooling}) at $r=R$ or the system becomes radiative and the adiabatic solution for $r>R$ given by CC85 is invalidated (see \citealt{wang,Silich03,Silich04, Thompson16}).
\begin{table*}
\begin{center}
\begin{tabular}{|l|l|l|r|r|r|}
\hline
Notation & Definition & Section/Eq. \\
\hline
 $\alpha$ & dimensionless thermalization efficiency & Section \ref{Sec_Anal}, eq.(\ref{parameter1})\\
 $\beta$  & dimensionless mass-loading efficiency & Section \ref{Sec_Anal}, eq.(\ref{parameter2})\\
 $R$ & starburst region & Section \ref{Sec_Anal}, eq.(\ref{winddensity})\\
 SFR & star formation rate in the galaxy & Section \ref{Sec_Anal}, eq.(\ref{winddensity}) \\
 $t_{\rm dyn}$ & dynamical timescale of the hot flow & Section \ref{Sec_Initial}, eq.(\ref{dyntime}) \\
 $t_{\rm cool}$ & cooling tiemscale of the hot flow & Section \ref{Sec_Initial}, eq.(\ref{cooltime})\\
 $T_c$ & temperature of the cloud & Section \ref{Sec_Initial}, eq.(\ref{pressure1})\\
 $n_{\rm H}^{\rm i}$ & initial hydrogen number density of the cloud & Section \ref{Sec_Initial}, eq.(\ref{pressure1})\\
 $r_0$ & starting position of the cloud & Section \ref{Sec_Initial} \\
 $\rho_0$,$u_0$,$P_0$ & dimensionless velocity, density and pressure of the hot wind at $r_0$ & Section \ref{Sec_Initial}, eq.(\ref{pressure2})\\
 $t_{\rm cc}$ & crushing tiemsacle of the cloud & Section \ref{Sec_Initial}, eq.(\ref{cctime}) \\
 $N_{\rm H}^{\rm i}$ & initial hydrogen column density of the cloud & Section \ref{Sec_Initial}, eq.(\ref{cctime})\\
 $R_c$ & radius of the cloud & Section \ref{Sec_Initial}, eq.(\ref{cctime}) \\
 $t_{\rm expan}$ & expansion timescale of the cloud & Section \ref{Sec_Initial}, eq.(\ref{expantime})\\
 $t_{\rm acc}$ & acceleration timescale of the cloud & Section \ref{Sec_Initial}, eq.(\ref{acctime1})\\
 $M_{\rm h}$ & Mach number of the hot flow around cloud & Section \ref{Sec_PE}, eq.(\ref{shtime1})\\
 $t_{\rm sh}$ & shredding timescale of the cloud & Section \ref{Sec_PE}, eq.(\ref{shtime})\\
 $\kappa$ & parameter in the cloud shredding timescale & Section \ref{Sec_PE}, eq.(\ref{shtime})\\
 $V_{c}^{\rm sh}$ & maximum velocity of the cloud estimated by $t_{\rm sh}$ & Section \ref{Sec_PE}, eq.(\ref{Vcsh})\\
 $t_{\rm evap}$ & evaporation timescale of the cloud & Section \ref{Sec_PE}, eq.(\ref{evaptime})\\
 $V_{c}^{\rm evap}$ & maximum velocity of the cloud estimated by $t_{\rm evap}$ & Section \ref{Sec_PE}, eq.(\ref{Vcevap})\\
 $\sigma$ & velocity dispersion of the galaxy & Section \ref{Sec_Gravity}, eq.(\ref{isothermal})\\
 $n_{\rm H}^{c}$ & hydrogen number density of cloud at pressure equilibrium with hot flow & Section \ref{Sec_Gravity}, eq.(\ref{balance1})\\
 $N_{\rm H}^{c}$ & hydrogen column density of cloud at pressure equilibrium with hot flow & Section \ref{Sec_Gravity}, eq.(\ref{balance2})\\
 $R_{c}^{\parallel}$, $R_{c}^{\perp}$ & cloud radius parallel and perpendicular to the hot flow & Section \ref{Sec_Gravity}, eq.(\ref{balance2})\\
 $\xi$ & $R_{c}^{\parallel}/ R_{c}^{\perp}$ & Section \ref{Sec_Gravity}, eq.(\ref{balance2})\\
 $\kappa_{\rm crit}$ & critical value of $\kappa$ that gives $t_{\rm sh}\sim t_{\rm acc}$ & Section \ref{Sec_kappa}, eq.(\ref{kappacrit})\\
\hline
\end{tabular}
\caption{Notations and definitions of some quantities in this paper.}\label{definitions}
\end{center}
\end{table*}


\begin{table*}
\begin{center}
Properties of CC85 Wind Solutions\\
\begin{tabular}{l|llll}
\hline
 & Eq. & $r_{0,*}=1$ & $r_{0,*}=2$ & $r_{0,*}=3$ \\
\hline
$u_{0}$    & & 0.71 & 1.26 & 1.33 \\
$\rho_{0}$ & & 0.11 & 1.58e-2 & 6.64e-3 \\
$P_{0}$ & & 3.37e-2 & 1.27e-3 & 3.02e-4 \\
$M_{\rm h}$ & & 1.0 & 3.44 & 4.84 \\
$(1+M_{\rm h})^{-1/6}\rho_{0}^{-1/2}u_{0}^{-1}P_{0}^{-1/2}$ & eq. (\ref{shtime}) & 20.4 & 137.8 & 395.3 \\
$(1+M_{\rm h})^{1/3}P_{0}\rho_{0}^{-1}$ & eq. (\ref{constraintsh}) & 0.38 & 0.13 & 8.19e-2 \\
$(1+M_{\rm h})^{-1/3}P_{0}^{-1}$ & eq. (\ref{flyingdistsh}) & 23.6 & 477.7 & 1840 \\
$(1+M_{\rm h})^{-1/3}M_{\rm h}^{-1/2}P_{0}^{-3/4}u_{0}^{-1}\rho_{0}^{-1/4}$ & eq. (\ref{evaptime}) & 24.6 & 108.7 & 290.0 \\
$(1+M_{\rm h})^{4/9}M_{\rm h}^{2/3}P_{0}\rho_{0}^{-1}$ &  eq. (\ref{constraintevap}) & 0.41 & 0.36 & 0.28 \\
$M_{\rm h}^{-1/2}(1+M_{\rm h})^{-1/3}P_{0}^{-3/4}u_{0}\rho_{0}^{3/4}$ & eq. (\ref{Vcevap}) & 1.39 & 2.73 & 3.41 \\
$M_{\rm h}^{-1}(1+M_{\rm h})^{-2/3}P_{0}^{-3/2}\rho_{0}^{1/2}$ & eq. (\ref{flyingdistevap}) & 34.1 & 297.0 & 990.5 \\
$(1+M_{\rm h})^{2/3}M_{\rm h}(u_{0}/r_{0,*})P_{0}^{3/2}\rho_{0}^{-3/2}$ & eq. (\ref{constraintevap2}) & 0.18 & 0.14 & 6.75e-2 \\
$(1+M_{\rm h})^{4}M_{\rm h}^{-8}\rho_{0}u_{0}^{-1}$ & eq. (\ref{constraintgrav0}) & 2.52 & 2.47e-4 & 1.93e-5 \\
$(1+M_{\rm h})^{-2/3}M_{\rm h}u_{0}^{1/2}P_{0}^{-1/6}$ & eq. (\ref{NHconstraint1}) & 0.93 & 4.35 & 6.65 \\
$(1+M_{\rm h})^{-2/3}P_{0}^{-2/3}$ & eq. (\ref{NHconstraint2}) & 6.04 & 31.5 & 68.5 \\
$(1+M_{\rm h})^{1/6}P_{0}^{1/2}\rho_{0}^{-1/2}$ & eq. (\ref{kappacrit}) & 0.61 & 0.36 & 0.29 \\
\hline
\end{tabular}
\caption{Here $r_{0,*}=r_{0}/R$, where $r_{0}$ is the the starting position of the cloud.}\label{CC85solutions}
\end{center}
\end{table*}

\cite{SH09} showed that for an axisymmetric disklike starburst, there is a spherical starburst CC85 model with an equivalent radius $R$ that can provide a good approximation in describing the hot wind solution in a disk-like starburst. The equivalent radius $R$ in general is smaller than the radius of the star forming disk region $R_{d}$. In the following sections we first take a fiducial value of $R=200$\,pc as the equivalent radius of galaxies for simplicity. Different radii are explored below.

It is believed that the pressure of the hot wind fluid will entrain cool gas clouds from the ISM (e.g., \citealt{Veilleux09}). In general, we expect the ISM of rapidly star-forming galaxies to be highly turbulent, with a broad lognormal distribution of densities and column densities and with a multi-phase medium. In order to explore constraints on the survival and dynamics of clouds, we first need to specify their properties. There are several parameters in our model for clouds: the temperature in the cloud $T_{c}$, the initial density and column density of the cloud $n_{\rm H}^{\rm i}$ and $N_{\rm H}^{\rm i}$, and the starting position (launching radius) of the cloud $r_{0}$. For simplicity, in our analytic estimates below and in Section \ref{Sec_Num}, we consider isothermal clouds with $T_c=10^2$, $10^3$, or $10^4$\,K as might be appropriate for molecular, neutral atomic, and ionized gas, respectively. Also, we consider clouds initially at radii $r\geq R$, and take $r_0=R$, $2R$ and $3R$. In general the parameters are scaled in terms of fiducial values $T_{c,3}=T_c/10^3$\,K,  $N_{\rm H}^{\rm i}=10^{21}N_{\rm H,21}^{\rm i}$ cm$^{-2}$, and $n_{\rm H}^{\rm i}=10^{3}n_{\rm H,3}^{\rm i}$ cm$^{-3}$.

Given these sets of parameters for the cool clouds, we now estimate the timecales that describe their dynamics and survival in a hot CC85-like flow, including the timescales for cloud crushing, expansion, acceleration, evaporation, and hydrodynamical instability (e.g., the Kelvin-Helmholz and Rayleigh-Taylor timescales). The pressure in the cloud is
\begin{equation}
P_{c}=1.4\times10^{-10}\;\textrm{dynes}\;\textrm{cm}^{-2}\;n_{\rm H,3}^{\rm i}T_{c,3}\label{pressure1}
\end{equation}
The pressure in the hot wind is $P_{\rm hot}=P_{\rm th}+P_{\rm ram}$, where $P_{\rm th}$ is the thermal pressure $k_{B}\rho_{\rm hot}T_{\rm hot}/(\mu m_{\rm H})$, and
and the ram pressure of the hot wind $P_{\rm ram}=\rho_{\rm hot}V_{\rm hot}^{2}$ is given by
\begin{eqnarray}
P_{\rm ram}=1.2\times10^{-7}\;\textrm{dynes}\;\textrm{cm}^{-2}\;\rho_{0}u_{0}^{2}\alpha^{1/2}\beta^{1/2}R_{200
\rm pc}^{-2}\textrm{SFR}_{1},\label{pressure2}
\end{eqnarray}
where the dimensionless velocity $u_{0}$, density $\rho_{0}$ are functions of radius (see Table \ref{CC85solutions}). Here $u_{0}$, $\rho_{0}$ and $P_{0}$ are $u_{*}$, $\rho_{*}$ and $P_{*}$ (see equations~\ref{WindT}, \ref{winddensity}, \ref{Vhot}) evaluated at $r_{*}=r_{0}$, respectively. Since the hot wind is supersonic (Mach number $M_{\rm h}\gg1$ for $r>R$), we have $P_{\rm ram}\gg P_{\rm th}$ and $P_{\rm hot}\simeq P_{\rm ram}$. If $P_{\rm ram}>P_{c}$, a shock will be driven into the cool clouds on a cloud crushing time when the hot wind overtakes the cool cloud, where the crushing time is defined as the time needed for the shock to cross the cloud:
\begin{eqnarray}
t_{\rm cc}&\approx&\frac{R_{c}}{v_{s}} \approx \frac{R_{c}}{V_{\rm hot}}\left(\frac{\rho_{c}}{\rho_{\rm hot}}\right)^{1/2}\nonumber\\
&\approx&2.2\times10^{3}\;\textrm{yr}\;\rho_{0}^{-1/2}u_{0}^{-1}\alpha^{-1/4}\beta^{-1/4}\nonumber\\
&&\times (n_{\rm H,3}^{i})^{-1/2} N_{\rm H,21}^{\rm i}R_{200\rm pc}\textrm{SFR}_{1}^{-1/2},\label{cctime}
\end{eqnarray}
and the shock velocity $v_{s}$ is estimated as $v_{s}=(P_{\rm ram}/\rho_{c})^{1/2}$ (\citealt{Klein94}; \citealt{Murray07}). On the other hand, if $P_{\rm ram}<P_{c}$, the cool gas cannot be pressure confined by the hot wind, and it will expand at its sound speed until the cloud reaches pressure equilibrium with the surrounding medium. Thus, the expansion timescale for pressure equilibrium is
\begin{eqnarray}
t_{\rm expan}&&\approx R_{c}\sqrt{\frac{m_{\rm H}}{k_{B}T_{c}}}\nonumber\\
&&\approx5.5\times10^{4}\;\textrm{yr}\;N_{\rm H,22}^{\rm i}(n_{\rm H,3}^{\rm i})^{-1}T_{c,3}^{-1/2}.\label{expantime}
\end{eqnarray}

The initial acceleration timescale of the cloud, i.e., the time for the cloud to become comoving with the hot wind flow is of order \footnote{The acceleration timescale $t_{\rm acc}$ is also called the drag timescale $t_{\rm drag}$ (e.g., \citealt{Faucher12}). Strictly, cool clouds can never reach the velocity of the hot wind, since the ram pressure on clouds decreases to zero while $V_{c}\rightarrow V_{\rm hot}$. The acceleration of a cloud is $a_{\rm ram}=3(V_{\rm hot}-V_{c})^{2}\rho_{\rm hot}/(4\rho_{c}R_{c})$, thus the acceleration timescale is estimated by
\begin{equation}
t_{\rm acc}=\int_{0}^{V_{c}^{\rm upper}}\frac{4R_{c}}{3 V_{\rm hot}}\left(\frac{\rho_{c}}{\rho_{\rm hot}}\right)\frac{d(V_{c}/V_{\rm hot})}{(1-V_{c}/V_{\rm hot})^{2}},\label{tacc}
\end{equation}
which diverges if we integrate $V_{c}$ from 0 to $V_{\rm hot}$. Analytically we estimate the cool cloud velocity $V_{c}$ to change from 0 to $V_{\rm upper}=V_{\rm hot}/2$ at a same position $r$, which yields Equation (\ref{tacc}) for the time for the cloud to reach half of $V_{\rm hot}$, and we say $V_{c}\sim V_{\rm hot}$ in this case. \label{footnote1}}
\begin{eqnarray}
t_{\rm acc}&&\approx\frac{4R_{c}}{3V_{\rm hot}}\left(\frac{\rho_{c}}{\rho_{\rm hot}}\right)\nonumber\\
&&\approx3.0\times10^{4}\;\textrm{yr}\;(\rho_{0}^{-1}u_{0}^{-1})\beta^{-1}N_{\rm H,21}^{\rm i}R_{200 \rm pc}^{2}\textrm{SFR}_{1}^{-1}\label{acctime1}
\end{eqnarray}
Comparing equations (\ref{cctime}) and (\ref{acctime1}), we have that for
\begin{equation}
\beta\leq\;32\,\rho_{0}^{-2/3}\alpha^{1/3}(n_{\rm H,3}^{\rm i})^{2/3}R_{200 \rm pc}^{4/3}\textrm{SFR}_{1}^{-2/3}\label{constraintcrushing}
\end{equation}
the crushing time is less than the acceleration time $t_{\rm cc}\leq t_{\rm acc}$ with $P_{\rm ram}>P_{c}$. Similarly, by comparing equations (\ref{expantime}) and (\ref{acctime1}), we find that for
\begin{equation}
\beta\leq\;0.55(\rho_{0}^{-1}u_{0}^{-1})R_{200 \rm pc}^{2}n_{\rm H,3}^{\rm i}T_{c,3}^{1/2}\textrm{SFR}_{1}^{-1},\label{constraintexpand}
\end{equation}
$t_{\rm expan}\leq t_{\rm acc}$ with $P_{\rm ram}< P_{c}$. Note that we have treated the cloud as isothermal, because the shocked gas inside the cloud quickly cools to $10^{4}$ K or below on a timescale of $\sim$ 100\,yr, much shorter than the timescales we consider below (\citealt{Murray07}; \citealt{Fujita09}). We take the temperature of the cool cloud as a constant, but always include the $T_{c}$ scaling.

\begin{figure*}
\centerline{\includegraphics[width=18cm]{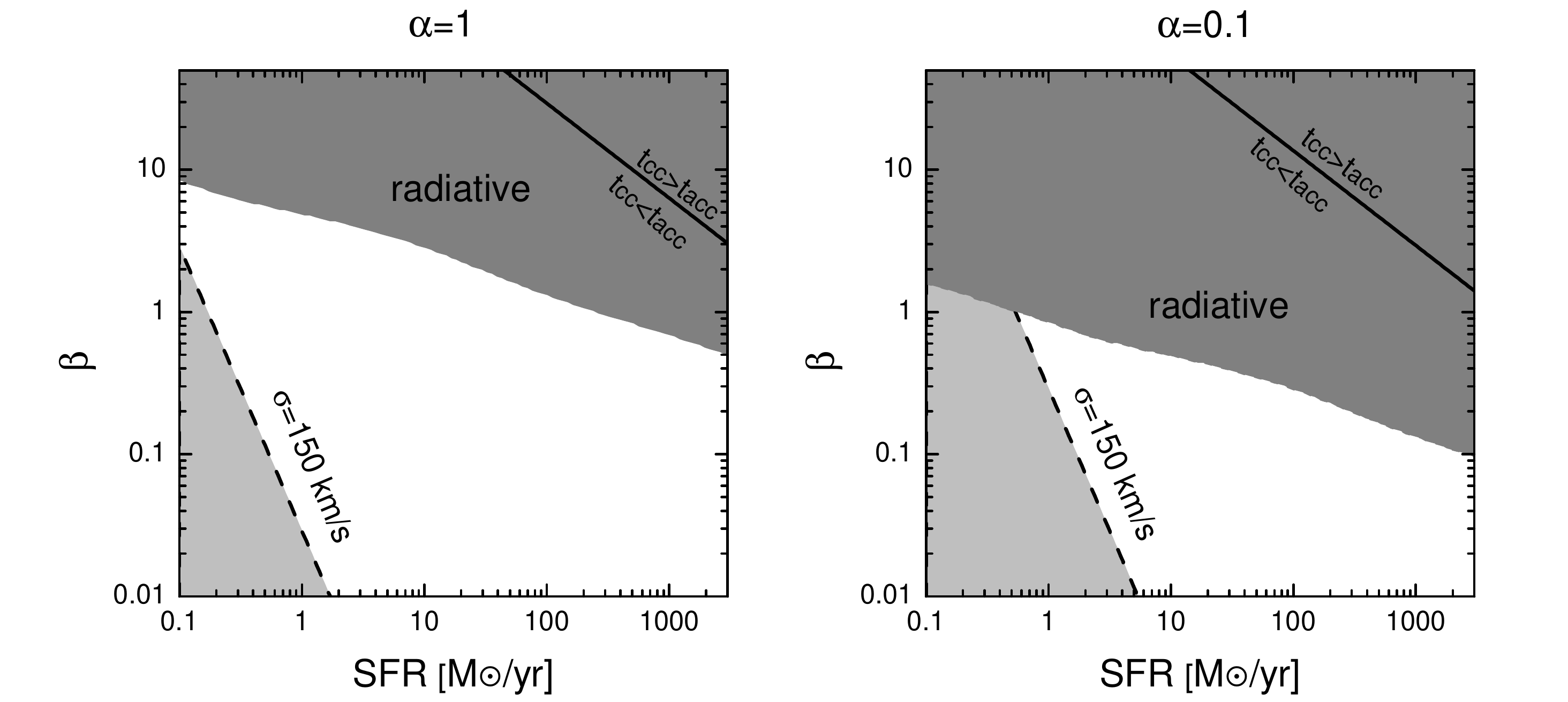}}
\caption{Timescale constraints and the gravity constraint as a function of SFR for clouds with $\alpha=1$ (left) and $\alpha=0.1$ (right), $T_{c}=10^{3}$ K, and the starting position $r_{0}=R$. Solid lines are the combined constraint of $t_{\rm cc}=t_{\rm acc}$ and $t_{\rm expan}=t_{\rm acc}$ (equations~\ref{constraintcrushing}, \ref{constraintexpand}). Dashed lines are the gravity constraint (equation~\ref{constraintgrav0}). The dark gray region shows where the flow is radiative (equation \ref{constraintcooling}), and the light gray region is excluded by the gravity constraint with $N_{\rm H}^{\rm i}=10^{20}$ cm$^{-2}$ and $\sigma=150$ km s$^{-1}$.}\label{fig_timeconstraint}
\end{figure*}

In Figure \ref{fig_timeconstraint} we show timescale constraints as a function of the mass loading efficiency $\beta$ and SFR for clouds with temperature $T_{c}=10^{3}$ K, taking the cloud starting position at $r_{0}=R$, and $\alpha=1$ (left) and $\alpha=0.1$ (right) as examples. The solid lines show the critical values of $\beta$ for $t_{\rm cc}=t_{\rm acc}$ in the case of $P_{\rm ram}>P_{c}$, or $t_{\rm expan}=t_{\rm acc}$ in the case of $P_{\rm ram}<P_{c}$. Over the regime plotted $P_{\rm ram} > P_{c}$, so only $t_{\rm cc}=t_{\rm acc}$ is shown. The dark gray regions are radiative, excluded by equation (\ref{constraintcooling}). Since the solid lines are always above the radiative cooling lines, for any hot flow with parameters in the non-radiative regime, the cool cloud will establish pressure equilibrium with the hot gas before being accelerated at $r_{0}=R$ with $T_{c}=10^{3}$ K. We find that this result is also valid for $T_{c}=10^{2}$ and $10^{4}$ K, and with varying $r_{0}$ from $R$ to $3R$. This means that in virtually all regimes of interest, clouds reach pressure equilibrium with the hot wind on a short timescale. We use this fact in the following analytical estimates. The dashed and dotted lines in Figure \ref{fig_timeconstraint} are discussed in Section \ref{Sec_Gravity}.

\subsection{Pressure Equilibrium and Cloud Destruction}\label{Sec_PE}

After pressure equilibrium with the hot flow, we can estimate whether cool clouds can be accelerated by ram pressure of the hot wind before being destroyed by hydrodynamical instabilities or thermal conduction and evaporation. We assume that after establishing pressure equilibrium at $r_{0}$, the cloud maintains pressure balance with the hot flow as it is accelerated. Although the pressure is strongest at the front of the cloud, and proportional to $P_{\rm hot}(1+M_{\rm h}^{2})$, \cite{SB15} showed that an oblique shock is formed at the extended cometary wind-cloud interface and that as a result the pressure equilibrium between the hot flow and the cool cloud is best described by $P_{c}\approx P_{\rm hot}(1+M_{\rm h})$. In our estimates below, we apply this scaling for $P_{c}$ and show how the Mach number ($M_{\rm h}$) of cold-hot pressure equilibrium enters the key expressions\footnote{Note that if one were to employ $P_{c}\approx P_{\rm hot}(1+M_{\rm h}^{2})$ for cloud pressure equilibrium, one finds higher pressures, smaller cloud radii, and more rapid destruction by hydrodynamical instabilities (equation~\ref{shtime}), leading to even smaller maximum cloud velocities (equations~\ref{Vcsh} and \ref{Vcevap}).}.

Simulations show that the shocked swept-up supershells in the central region of molecular disks quickly cool and fragment through Kelvin-Helmholtz (KH) or Rayleigh-Taylor (RT) instabilities (\citealt{SS00}; \citealt{Heckman00}; \citealt{Fujita09}), which have comparable timescales (\citealt{Krolik81}; \citealt{Schiano95}). \cite{Faucher12} (see also \citealt{Klein94}) suggested that the timescale for coulds to be destroyed by the KH instability is $t_{\rm KH}\approx 10\;t_{cc}^{\rm th}$, where $t_{cc}^{\rm th}$ is the crushing time of a cloud which is initially in thermal pressure equilibrium with the hot medium ($\rho_{c}^{\rm th}T_{c}=\rho_{\rm hot}T_{\rm hot}$). However, recent simulations show that the cloud destruction timescale  may depend on the Mach number of the flow. In particular, \cite{SB15} showed that clouds are destroyed by the KH instability only after they are shredded by other hydrodynamical instabilities. They found that the timescale for 50\% of cloud to be below $2/3$ of the initial cloud density is
\begin{equation}
t_{50}=4 t_{\rm cc}^{\rm th} \sqrt{1+M_{\rm h}}\label{shtime1}.
\end{equation}
\cite{SR16} did similar high-resolution simulations of cloud destruction for both turbulent and spherical clouds, and found a longer lifetime for spherical clouds. The difference is caused by the different treatment of cooling in the simulations. In \cite{SB15} the clouds only allow cooling above $Tc\gtrsim 10^{4}$ K with the assumption of complete ionization, but in \cite{SR16} the temperature of the post-shock gas can be down to $\sim 100$ K. Here, we follow \cite{SB15} and assume that a cloud is destroyed on the shredding timescale
\begin{eqnarray}
t_{\rm sh}&\approx&\kappa\left(\frac{\rho_c^{\rm th}}{\rho_{\rm hot}}\right)^{1/2}\frac{R_{c}^{\rm th}}{V_{\rm hot}}(1+M_{\rm h})^{1/2}\nonumber\\
&\approx&3.1\times10^{2}\;\textrm{yr}\;\kappa_{4}(1+M_{\rm h})^{-1/6}\rho_{0}^{-1/2}u_{0}^{-1}P_{0}^{-1/2}\nonumber\\
&&\times \alpha^{-1/2}\beta^{-1/2}N_{\rm H,21}^{c}T_{c,3}^{1/2}R_{200 \rm pc}^{2}\textrm{SFR}_{1}^{-1},\label{shtime}
\end{eqnarray}
where $\kappa_{4}=\kappa/4$ is a constant. After this time the cloud is considered to be destroyed.  We use $T_{c}=10^{3}T_{c,3}$ K as the fiducial value. A turbulent cloud can be considered to have a lower temperature or a lower $\kappa$, which gives a shorter timescale of $t_{\rm sh}$. Comparing the two timescales $t_{\rm sh}$ and $t_{\rm acc}$ in the case of pressure equilibrium, we find that if
\begin{equation}
\beta \leq 9.5\times10^{3}\,(1+M_{\rm h})^{1/3}P_{0}\rho_{0}^{-1}\alpha \kappa_{4}^{-2}T_{c,3}^{-1},\label{constraintsh}
\end{equation}
then $t_{\rm sh}\leq t_{\rm acc}$, and the cloud should be shredded before acceleration to $V_{\rm hot}$. Note that the factor $P_{0}/\rho_{0}$ strongly decreases with radius, such that $(1+M_{\rm h})^{1/3}P_{0}\rho_{0}^{-1}\simeq 0.4-0.08$ for $r_{0}=R$ to $3R$. Since $\beta\lesssim 1$ is required for hot winds from the X-ray constraints presented in \cite{Zhang14}, equation (\ref{constraintsh}) is a strong constraint. It implies that $t_{\rm sh}$ is essentially always longer than $t_{\rm acc}$ for $\kappa\lesssim390(1+M_{\rm h})^{1/6}P_{0}^{1/2}\rho_{0}^{-1/2}\alpha^{1/2}T_{c,3}^{-1/2}$. More discussion of larger $\kappa$ and its implications for our results is given in Section \ref{Sec_kappa}. In the $\beta-$SFR plane shown in the two panels of Figure \ref{fig_timeconstraint}, equation (\ref{constraintsh}) is a horizontal line off the top of both plots; for the fiducial model, $t_{\rm sh}$ is always much smaller than $t_{\rm acc}$.

We can then estimate the maximum velocity $V_{c}$, and the ``flying distance" $\Delta r$ of the cloud, i.e., the distance between the cloud starting position $r_{0}$ to its destruction position $r_{0}+\Delta r$, accelerated in a timescale of $t_{\rm sh}$ respectively. If the cloud is destroyed by the shredding timescale, we have
\begin{eqnarray}
V_{c}^{\rm sh}&=&a_{c}t_{\rm sh}=\frac{3}{4}\kappa V_{\rm hot}\left(\frac{\rho_{c}}{\rho_{\rm hot}}\right)^{-1/2}(1+M_{\rm h})^{1/3}\nonumber\\
&\approx&10\;\textrm{km}\;\textrm{s}^{-1}\;M_{\rm h}(1+M_{\rm h})^{-1/6}\kappa_{4}T_{c,3}^{1/2},\label{Vcsh}
\end{eqnarray}
\begin{eqnarray}
\Delta r_{\rm sh}&=&\frac{1}{2}a_{c}t_{\rm sh}^{2}=\frac{3}{8}\kappa^{2}R_{c}(1+M_{\rm h})^{2/3}\nonumber\\
&\approx&1.1\times10^{-3}\;\textrm{pc}\;P_{0}^{-1}(1+M_{\rm h})^{-1/3}\kappa_{4}^{2}\nonumber\\
&&\times \alpha^{-1/2}\beta^{-1/2}N_{\rm H,21}^{c}T_{c,3}R_{200 \rm pc}^{2}\textrm{SFR}_{1}^{-1},\label{flyingdistsh}
\end{eqnarray}
where $a_{c}$ is the acceleration of the cloud. Note that $V_{c}^{\rm sh}$ is only a function of $\kappa$ and $T_{c}$, and is always below $100$ km s$^{-1}$ for the fiducial model. For $r_{0}=R$ $(3R)$ with $R=200$ pc we have $\Delta r_{\rm sh}\approx 0.03(2)\;\textrm{pc}\;N_{\rm H,21}^{c}T_{c,3}\textrm{SFR}_{1}^{-1}$, as long as $\Delta r$ is small compared with $R$. These results show that the cloud is destroyed very near its starting position with a low velocity $V_{c}^{\rm sh}$, but with a strong dependence on $\kappa$. Because $M_{\rm h}\propto r^{2/3}$ in the CC85 model, the maximum velocity $V_{c}^{\rm sh}\propto M_{\rm h}^{5/6}\propto r^{5/9}$, thus the starting position of cloud is important, especially for large starting position $r_{0}$ (see Section \ref{Sec_Fiducial}).

Thermal conduction may also be important to evaporate the clouds (e.g., \citealt{Cowie77}; \citealt{Krolik81}; \citealt{BS16}). Following \cite{BS16}, we adopt the timescale for cloud evaporation
\begin{equation}
t_{\rm evap} \approx \frac{100}{f(M_{\rm h})}\left(\frac{\rho_{c}^{\rm th}}{\rho_{\rm hot}}\right)^{-1/2}\frac{2g}{\sqrt{1+4g}-1},
\end{equation}
where the functions $f(M_{\rm h})$ and $g$ are given in \cite{BS16} (see their equations 11 and 19). Using the cloud and hot flow parameters we find that
\begin{eqnarray}
t_{\rm evap}&\approx&22\;\textrm{yr}\;(1+M_{\rm h})^{-1/3}M_{\rm h}^{-1/2}P_{0}^{-3/4}u_{0}^{-1}\rho_{0}^{-1/4}\nonumber\\
&&\times \alpha^{-3/4}\beta^{-1/4}(N_{\rm H,21}^{c})^{1/2}T_{c,3}^{3/4}R_{200 \rm pc}^{2}\textrm{SFR}_{1}^{-1}.\label{evaptime}
\end{eqnarray}
Setting $t_{\rm evap}\leq t_{\rm acc}$ implies an upper limit on $\beta$ of
\begin{eqnarray}
\beta\leq 1.5\times 10^{4}\,(1+M_{\rm h})^{4/9}M_{\rm h}^{2/3}P_{0}\rho_{0}^{-1}\alpha(N_{\rm H,21}^{c})^{2/3}T_{c,3}^{-1}.\label{constraintevap}
\end{eqnarray}
The dimensionless factor $(1+M_{\rm h})^{4/9}M_{\rm h}^{2/3}P_{0}\rho_{0}^{-1}\sim0.41-0.28$ from $r_{0}=R$ to $3R$ (see Table \ref{CC85solutions}). The constraint on $\beta$ given by equation (\ref{constraintevap}) is always stronger than the constraint given in equation (\ref{constraintsh}) for the cloud shredding timescale unless $N_{\rm H}^{c}\leq 5\times10^{20}$ cm$^{-2}$ $(1+M_{\rm h})^{-1/6} M_{\rm h}^{-1}\kappa_{4}^{-2}$. In Section \ref{Sec_Gravity} we combine a constraint on $N_{\rm H}^{c}$ derived by comparing the gravitational and ram pressure forces on clouds, and show that equation (\ref{constraintevap}) always holds for cool clouds accelerated outwards by adiabatic hot winds.

The evaporation may play an important role in destroying the cloud. The maximum velocity of the cloud if it is subject to only evaporation is
\begin{eqnarray}
V_{c}^{\rm evap}&=&a_{c}t_{\rm evap}=\frac{3V_{\rm hot}^{2}}{4R_{c}}\left(\frac{\rho_{\rm hot}}{\rho_{c}}\right)t_{\rm evap}\nonumber\\
&\approx&0.6\;\textrm{km}\;\textrm{s}^{-1}\;M_{\rm h}^{-1/2}(1+M_{\rm h})^{-1/3}P_{0}^{-3/4}u_{0}\rho_{0}^{3/4}\nonumber\\
&& \times\alpha^{-1/4}\beta^{1/4}(N_{\rm H,21}^{c})^{-1/2}T_{c,3}^{3/4},\label{Vcevap}
\end{eqnarray}
which is significant lower than $V_{c}^{\rm sh}$ given by equation (\ref{Vcsh}). The distance traveled before destruction is
\begin{eqnarray}
\Delta r_{\rm evap}&=&\frac{1}{2}a_{c}t_{\rm evap}^{2}\nonumber\\
&\approx&7\times10^{-5}\;\textrm{pc}\;M_{\rm h}^{-1}(1+M_{\rm h})^{-2/3}P_{0}^{-3/2}\rho_{0}^{1/2}\nonumber\\
&&\times \alpha^{-1}T_{c,3}^{3/2}R_{200 \rm pc}^{2}\textrm{SFR}_{1}^{-1},\label{flyingdistevap}
\end{eqnarray}
These conclusions of low $V_{c}$ and small $\delta r$ are essentially similar as for the cloud destroyed by shredding. If thermal condition is important, the cloud is even more difficult to be accelerated than the non-conduction case. 

\begin{figure}
\centerline{\includegraphics[width=10cm]{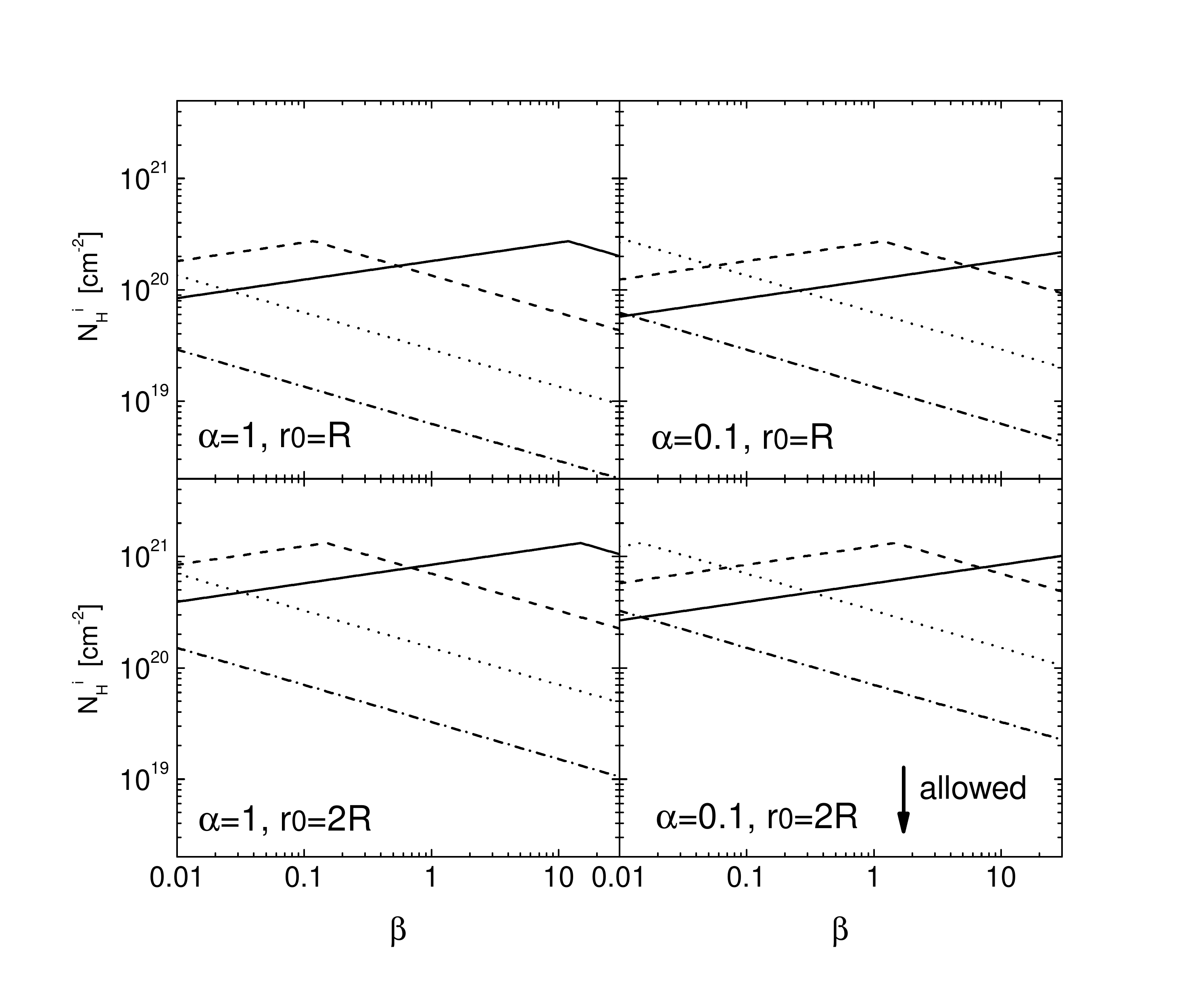}}
\caption{Constraint on initial cloud column density $N_{\rm H}^{\rm i}$ as a function of $\beta$ for $\alpha=1$ (left panels) and $\alpha=0.1$ (right panels) and for $r_{0}=R$ (upper panels) and $r_{0}=2R$ (lower panels), for $R=200$ pc, SFR $=1\;M_{\odot}$ yr$^{-1}$ (solid), $10\;M_{\odot}$ yr$^{-1}$ (dashed), $100\;M_{\odot}$ yr$^{-1}$ (dotted), $1000\;M_{\odot}$ yr$^{-1}$ (dash-dotted) and taking $N_{\rm H}^{\rm obs}=10^{21}$ cm$^{-2}$ (see equations \ref{NHconstraint1} and \ref{NHconstraint2}), $\sigma=150$ km s$^{-1}$.}\label{fig_NHconstraint}
\end{figure}

\subsection{Constraints on Cloud Column Density from Gravity}\label{Sec_Gravity}


In the case of $V_{\rm hot}\gg V_{c}$, where $V_{c}$ is the velocity of the cool cloud, the ram pressure force at the front of the cloud is $F_{\rm ram}\approx\rho_{\rm hot}V_{\rm hot}^{2}A_{c}$, where $A_{c}=\pi R_{c}^{2}$ is the projected area of the cloud. In order for the cloud to be accelerated by the hot flow, the ram pressure must be stronger than gravity after pressure equilibrium is established. For simplicity if we take an isothermal sphere model for the gravitational potential of the galaxy with $M_{\rm gal}(r)=2\sigma^{2}r/G$, where $\sigma$ is the velocity dispersion of the galaxy, the gravitational force is
\begin{equation}
F_{\rm grav}=2\sigma^{2}M_{c}/r,\label{isothermal}
\end{equation}
where $M_{c}=4\pi \rho_{c}R_{c}^{3}/3$ is the total mass of the cloud. The requirement $F_{\rm ram}>F_{\rm grav}$ gives a constraint on the column density of the cloud after pressure equilibrium of
\begin{eqnarray}
N_{\rm H,21}^{c}\leq\;8.2\;(u_{0}r_{0,*}^{-1})\alpha^{1/2}\beta^{1/2}\sigma_{150}^{-2}R_{200 \rm pc}^{-1}\textrm{SFR}_{1},\label{NHc_constraint1}
\end{eqnarray}
where $r_{0,*}=r_{0}/R$ is the dimensionless radial starting position of the cloud (see Table \ref{CC85solutions}). Combining equation (\ref{NHc_constraint1}) with (\ref{constraintevap}) to eliminate the column density dependence, we find that the constraint on $\beta$ for $t_{\rm evap}\leq t_{\rm acc}$ is
\begin{eqnarray}
\beta&\leq& 1.5\times10^{7}\,\left(\frac{u_{0}}{r_{0,*}}\right)(1+M_{\rm h})^{2/3}M_{\rm h}P_{0}^{3/2}\rho_{0}^{-3/2}\nonumber\\
&&\times\alpha^{2}\sigma_{150}^{-2}T_{c,3}^{-3/2}R_{200 \rm pc}^{-1}\textrm{SFR}_{1}.\label{constraintevap2}
\end{eqnarray}
We find that the constraint of $\beta$ given by equation (\ref{constraintevap2}) is always in the radiative region of the (SFR$,\beta)$ parameter space, which means that the cloud will always be destroyed before being accelerated for non-radiative hot winds.


For simplicity, if we assume the initial cloud is compressed by the ram pressure of the hot wind in a timescale of $t_{\rm cc}$ (equation \ref{cctime}) and comes into pressure equilibrium with the hot wind, we can relate the hydrogen density and column density of the cloud after pressure equilibrium ($n_{\rm H}^{c}$ and $N_{\rm H}^{c}$ ) to its initial values ($n_{\rm H}^{\rm i}$ and $N_{\rm H}^{\rm i}$):
\begin{equation}
n_{\rm H}^{c}=8.5\times10^{5}\;\textrm{cm}^{-3}\;(1+M_{\rm h})P_{0}\alpha^{1/2}\beta^{1/2}R_{200 \rm pc}^{-2}T_{c,3}^{-1}\textrm{SFR}_{1},\label{balance1}
\end{equation}
and
\begin{eqnarray}
N_{\rm H}^{c}&=&90\;N_{\rm H}^{\rm i}\xi^{2/3}(1+M_{\rm h})^{2/3}P_{0}^{2/3}\alpha^{1/3}\beta^{1/3}R_{200\rm pc}^{-4/3}\nonumber\\
&&\times (n_{\rm H,3}^{\rm i})^{-2/3}T_{c,3}^{-2/3}\textrm{SFR}^{2/3}_{1}\label{balance2}
\end{eqnarray}
respectively. Simulations shows that the compression of the cloud is almost completely perpendicular to the hot flow, thus we introduce a factor $\xi = R_{c}^{\parallel}/R_{c}^{\perp}$ in equation (\ref{balance2}), where $R_{c}^{\parallel}$ and $R_{c}^{\perp}$ are the radius of the cloud parallel and perpendicular to the flow respectively. Typically in the simulations of \cite{SB15}, $R_{c}^{\parallel}/R_{c}^{\perp}\sim 8$ on a timesacle of $t_{\rm sh}$.

Using Equations (\ref{balance1}) and (\ref{balance2}), the Eddington-like limit given by the constraint $F_{\rm ram}\geq F_{\rm grav}$ then translates into a constraint on $\beta$:
\begin{eqnarray}
\beta&\geq&113\,(1+M_{\rm h})^{4}M_{\rm h}^{-8}\rho_{0}u_{0}^{-1}\alpha^{-1}\xi^{4}(N_{\rm H,21}^{\rm i})^{6}(n_{\rm H,3}^{\rm i})^{-4}\nonumber\\
&&\times T_{c,4}^{-3}\sigma_{150}^{12}R_{200 \rm pc}^{-2}\textrm{SFR}_{1}^{-2}.\label{constraintgrav0}
\end{eqnarray}
For simplicity we take $\xi=1$, which gives a lower limit on the {\it minimum} value of $\beta$ required for acceleration. The dotted lines in Figure \ref{fig_timeconstraint} show this limit at $r_{0}=R$ for $\sigma=150$\,km s$^{-1}$ and clouds with initial $N_{\rm H}^{\rm i}=10^{20}$ cm$^{-2}$, where the light gray regions are excluded by equation (\ref{constraintgrav0}). Since the Eddington-like limit on $\beta$ is extremely sensitive to virtually all of the parameters of the problem ($\beta\propto (N_{\rm H}^{\rm i})^{6}$ in equation \ref{constraintgrav0}), clouds with initial $N_{\rm H}^{\rm i}=10^{21}$ cm$^{-2}$ are unlikely to be accelerated at $r_{0}=R$ because of gravity. However, note that since the critical value of $\beta$ is so sensitive to the set of parameters, the gravity constraint at fixed $\sigma$ is weak. The strong $\xi$ and $N_{\rm H}$ dependence of $\beta$ in equation (\ref{constraintgrav0}) implies that simulations of cloud acceleration and destruction should be explored including the effects of gravity.

If we take $\alpha$, $\beta$ and $\sigma$ as fixed parameters, equation (\ref{constraintgrav0}) can be written as a constraint on the initial cloud column density $N_{\rm H}^{\rm i}$ such that $F_{\rm ram}\geq F_{\rm grav}$:
\begin{eqnarray}
N_{\rm H,21}^{\rm i}&\leq&\;0.41(1+M_{\rm h})^{-2/3}M_{\rm h}u_{0}^{1/2}P_{0}^{-1/6}\alpha^{1/6}\beta^{1/6}\nonumber\\
&&\times(n_{\rm H,3}^{\rm i})^{2/3}T_{c,3}^{2/3}\sigma_{150}^{-2}R_{200pc}^{1/3}\textrm{SFR}_{1}^{1/3},\label{NHconstraint1}
\end{eqnarray}
which gives an upper bound on the initial cloud column density $N_{\rm H}^{\rm i}$ for ejection from a galaxy, where the dimensionless factor $(1+M_{\rm h})^{-2/3}M_{\rm h}u_{0}^{1/2}P_{0}^{-1/6}$ increases from $\sim0.9$ to $6.6$ from $r_{0}=R$ to $r_{0}=3R$ (Table \ref{CC85solutions}).

On the other hand, $N_{\rm H}^{\rm i}$ can be constrained by observations. The measured Na D or Mg II column density in the outflows of LIRGs and ULIRGs gives a constraint on the observationally-derived total hydrogen column density of $N_{\rm H}^{\rm obs}\sim10^{20}-10^{21}$ cm$^{-2}$, with an order of magnitude uncertainty due to the metallicity of the gas, the Na depletion factor, and the Na ionization correction (e.g., \citealt{Heckman00}; \citealt{Schwartz04}; \citealt{Rupke02}; \citealt{Rupke05b}; \citealt{Martin05}; \citealt{Martin06}; \citealt{Murray07}). It has been shown that the atomic absorption lines are optically thick, with a typical covering factor of $C_{f}\sim0.2-1$. Assuming the apparent column density of the cloud obtained by observation is $N_{\rm H}^{\rm obs}$, with an amplification factor of $C_{f}^{-1}$, the total column density along the line of sight $N_{\rm H}^{\rm obs}C_{f}^{-1}$ is contributed to by multiple overlapping single clouds with a column density of $N_{\rm H}^{c}$, thus we have $N_{\rm H}^{c}\leq N_{\rm H}^{\rm obs}C_{f}^{-1}$, which gives
\begin{eqnarray}
N_{\rm H,21}^{\rm i}&\leq&0.01(1+M_{\rm h})^{-2/3}P_{0}^{-2/3}\alpha^{-1/3}\beta^{-1/3}N_{\rm H,21}^{\rm obs}C_{f}^{-1}\nonumber\\
&&\times (n_{\rm H,3}^{\rm i})^{2/3}R_{200 \rm pc}^{4/3}T_{c,3}^{2/3}\textrm{SFR}_{1}^{-2/3}.\label{NHconstraint2}
\end{eqnarray}

Figure \ref{fig_NHconstraint} demonstrates examples on the upper bounds on $N_{\rm H}^{\rm i}$ as a function of $\beta$ for various SFR and $\alpha$, where we choose a typical value for $N_{\rm H}^{\rm obs}=10^{21}$ cm$^{-2}$, a covering factor of $C_{f}=0.5$, and $T_{c}=10^{3}$ K, $n_{\rm H}^{\rm i}=10^{3}$ cm$^{-3}$, $R=200$ pc and $\sigma=150$ km s$^{-1}$ in equations (\ref{NHconstraint1}) and (\ref{NHconstraint2}). Higher SFR yields a more stringent constraint on $\beta$ and $N_{\rm H}^{\rm i}$. The constraint on $N_{\rm H}^{\rm i}$ is weaker for larger initial cloud launch radius $r_{0}$. For example, $N_{\rm H}^{\rm i}$ is always $N_{\rm H}^{\rm i}\lesssim 3\times10^{20}$\,cm$^{-2}$ at $r_{0}=R$. For ${\rm SFR}\gtrsim 100$\,M$_{\odot}$ yr$^{-1}$, $N_{\rm H}^{\rm i}\lesssim 2\times10^{20}$\,cm$^{-2}$ at $r_{0}=R$ and $N_{\rm H}^{\rm i}\lesssim 10^{21}$\,cm$^{-2}$ at $r_{0}=2R$. Note that the constraint of $N_{\rm H}^{\rm i}\lesssim10^{20}-10^{21}$ cm$^{-2}$ is given for the fiducial parameter set. Higher values of $T_{c}$, $n_{\rm H}^{\rm i}$, or $R$ can increase the upper bound on $N_{\rm H}^{\rm i}$. For example, for $T_{c}=10^{4}$ K and $R=1$ kpc, we derive that $N_{\rm H}^{\rm i}\lesssim 10^{22}$ cm$^{-2}$.

However, note that because both $N_{\rm H}^{\rm obs}$ and $C_{f}$ have an order of magnitude uncertainty, the constraint given by Equation (\ref{NHconstraint2}) has at least one order of magnitude uncertainty. Even so, we find that, in general $N_{\rm H}^{\rm i}$ is constrained to be $N_{\rm H}^{\rm i}\lesssim 10^{20}-10^{22}$ cm$^{-2}\;(n_{\rm H}^{\rm i})^{2/3}T_{c}^{2/3}$ for $R\geq 200$\,pc over a broad range of SFR.

\begin{figure*}
\begin{center}
\centerline{\includegraphics[width=18cm]{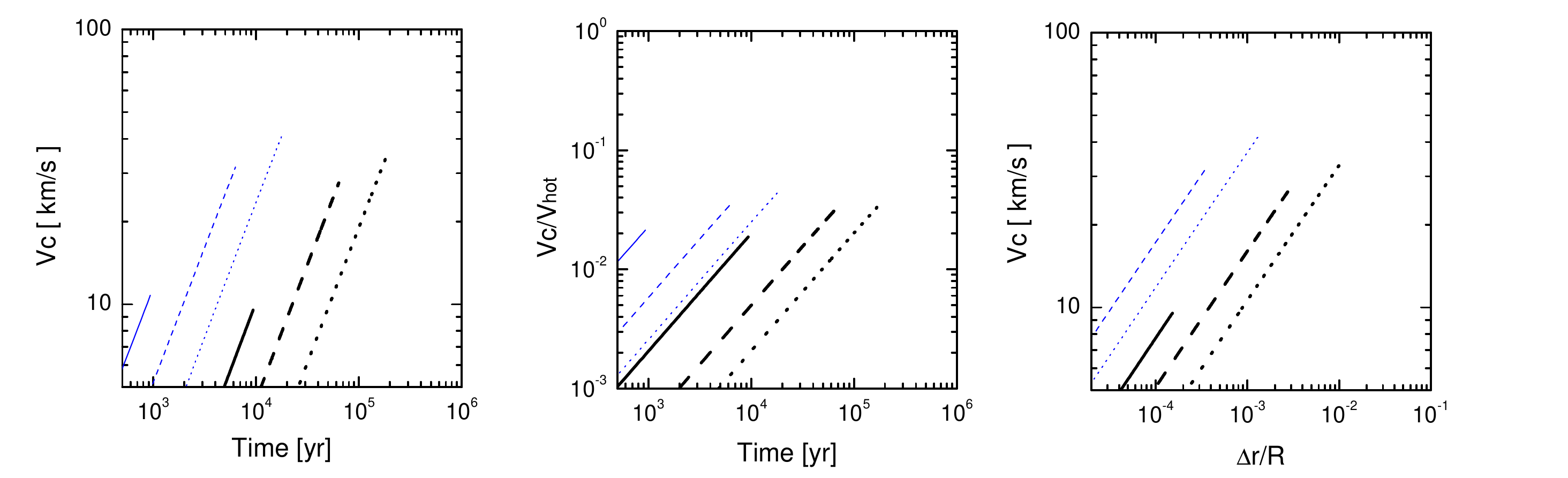}}
\centerline{\includegraphics[width=18cm]{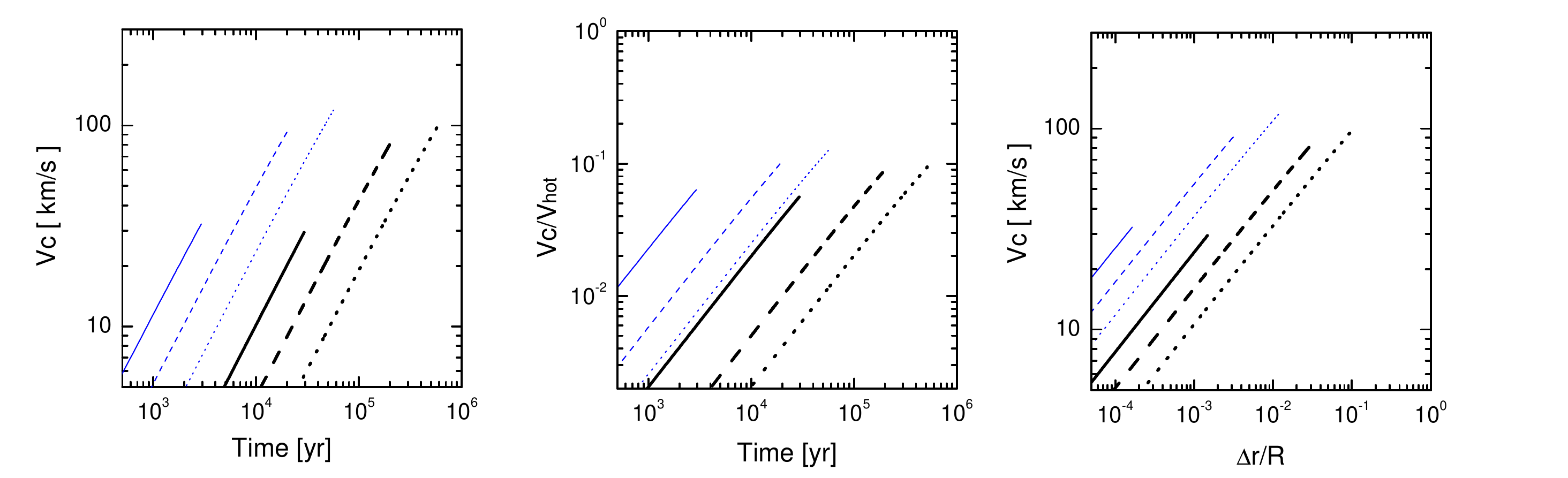}}
\caption{Velocity of a cool cloud $V_{c}$ and $V_{c}/V_{\rm hot}$ as functions of time (left two panels), and $V_{c}$ as a function of radius $\Delta r/R$(right panels) with cloud temperature $T_{c}=10^{3}$ K (upper panels), $T_{c}=10^{4}$ K (lower panels), and $N_{\rm H}^{c}=10^{21}$ cm$^{-2}$ (black thick lines), $N_{\rm H}^{c}=10^{20}$ cm$^{-2}$ (blue thin lines), and starting position at $r_{0}=R$ (solid lines), $2R$ (dashed lines), $3R$ (dotted lines), where $(\alpha,\beta)=(1,1)$, and the host galaxy has $R=200$ pc, SFR$=10\,M_{\odot}$ yr$^{-1}$, and $\sigma=150$ km s$^{-1}$.}\label{fig_Vc}
\end{center}
\end{figure*}
\begin{figure*}
\centerline{\includegraphics[width=18cm]{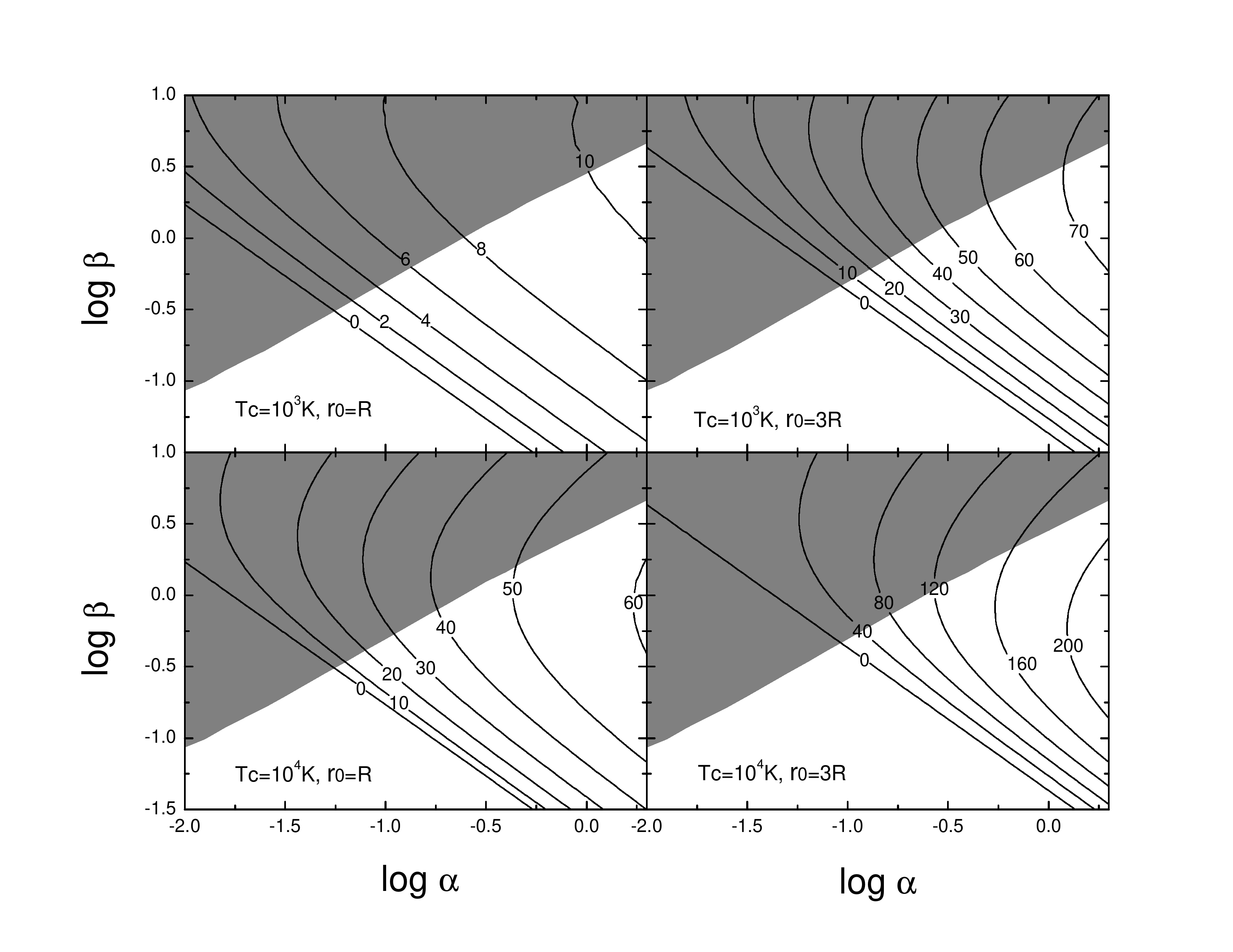}}
\caption{Contours of the maximum velocity of cool gas $V_{c}$ (km s$^{-1}$) in the parameter space of $(\log\alpha,\log\beta)$, with cloud starting position $r_{0}=R$ (left panels), $3R$ (right panels) with $R=200$ pc, $T_{c}=10^{3}$ (upper panels) and $T_{c}=10^{4}$ K (lower panels), where $N_{\rm H}^{c}=10^{21}$ cm$^{-2}$, and host galaxy SFR$=10\,M_{\odot}$ yr$^{-1}$ and $\sigma=150$ km s$^{-1}$. The shaded regions indicate that the solution is radiative at $R$. The contours go to $V_{c}=0$ in the lower left region of each panel because of the gravity constraint of equation (\ref{NHc_constraint1}).}\label{fig_VcContour}
\end{figure*}
\begin{figure*}
\centerline{\includegraphics[width=18cm]{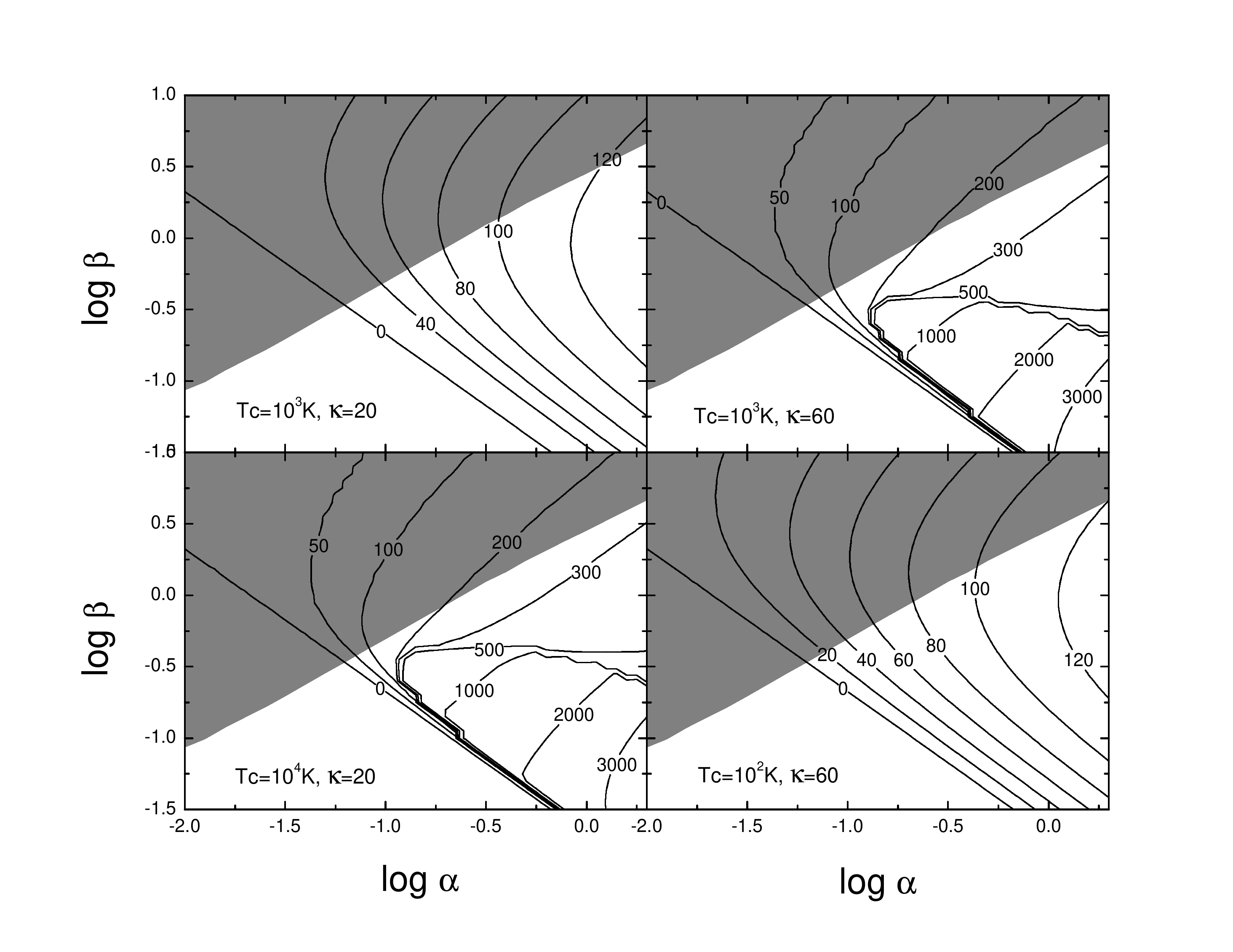}}
\caption{Contours of the maximum  velocity of cool gas $V_{c}$ (km s$^{-1}$) in the parameter space of $(\log\alpha,\log\beta)$ with various $T_{c}$ and $\kappa$: $\kappa=20$ (left panels), $\kappa=60$ (lower panels), $T_{c}=10^{3}$ K (upper panels), $T_{c}=10^{4}$ K (lower left), $T_{c}=10^{2}$ K (lower right), with cloud starting position $r_{0}=2R$, $R=200$ pc, SFR$=10\,M_{\odot}$ yr$^{-1}$, $\sigma=150$ km s$^{-1}$, and $N_{\rm H}^{c}=10^{21}$ cm$^{-2}$.}\label{fig_Vckappa}
\end{figure*}

\section{Numerical Solutions of the Ram Pressure Acceleration of Cool Clouds}\label{Sec_Num}

\subsection{Fiducial Model}\label{Sec_Fiducial}

In this section we calculate the cloud evolution numerically. Assuming that ram pressure dominates the driving of cool gas clouds, the equation of motion for a cloud of cool gas is
\begin{equation}
M_{c}\frac{dV_{c}}{dt}=\dot{M}_{\rm hot}V_{\rm hot}\left(1-\frac{V_{c}}{V_{\rm hot}}\right)^{2}\frac{A_{c}}{\Omega r^{2}}-\frac{GM_{\rm gal}(r)M_{c}}{r^{2}},\label{rpa1}
\end{equation}
where $\dot{M}_{\rm hot}$ is the mass-loss rate of the hot wind, $\Omega$ is the solid angle subtended by hot wind fluid, and $M_{\rm gal}(r)$ is the mass of the galaxy. For the spherical model $\Omega=4\pi$. The cloud radius $R_{c}$ evolves as a function of time in response to the cloud's internal pressure $P_{c}$ and the surrounding hot fluid. For $P_{\rm hot}(1+M_{\rm h})<P_{c}$ we use
\begin{equation}
\frac{dR_{c}}{dt}=\sqrt{\frac{k_{B}T_{c}}{m_{\rm H}}}\left[1-\frac{P_{\rm hot}(1+M_{\rm h})}{P_{c}}\right]^{1/2}\label{equRc1}
\end{equation}
On the other hand, if $P_{\rm hot}(1+M_{\rm h})>P_{c}$, the cloud is compressed. As mentioned in Section \ref{Sec_Gravity}, in this case the compression of the cloud is almost completely perpendicular to the hot flow, we use
\begin{equation}
\frac{dR_{c}^{\perp}}{dt}=-\sqrt{\frac{k_{B}T_{c}}{m_{\rm H}}}\left[\frac{P_{\rm hot}(1+M_{\rm h})}{P_{c}}-1\right]^{1/2}\label{equRc2}
\end{equation}
for $P_{\rm hot}(1+M_{\rm h})>P_{c}$, where $R_{c}^{\perp}$ is the radius of the cloud perpendicular to the flow. As in Sections \ref{Sec_PE} and \ref{Sec_Gravity},  we take the pressure on the cloud to be $P_{\rm hot}(1+M_{\rm h})$ when calculating pressure equilibrium with the hot gas, even though the ram pressure at the head of the cloud is proportional to $\rho_{\rm hot} V_{\rm hot}^2$, consistent with the numerical results of \cite{SB15}. For a given parameter set of $(\alpha,\beta)$, $r_{0}$, and SFR, the cloud velocity $V_{c}$ can be calculated by solving equations (\ref{rpa1}), (\ref{equRc1}) and (\ref{equRc2}). We require $t_{\rm sh}>t$ in the calculation, otherwise the cloud should be destroyed and the calculation stops. If $t_{\rm evap}$ is taken into account and $t_{\rm evap}<t_{\rm sh}$, we have an even more stringent constraint on the maximum velocities $V_{c}$.

We start by calculating the cloud evolution for the fiducial model ($\kappa=4$) for cloud destruction based on high-resolution hydrodynamical simulations (equation~\ref{shtime}), and compare with the analytical results in Section \ref{Sec_Anal}. Figure \ref{fig_Vc} gives examples of solutions for $V_{c}$, the ratio $V_{c}/V_{\rm hot}$ as functions of time, and $V_{c}$ as a function of radius for different cloud properties: $T_{c}=10^{3}$\,K (upper panels), $T_{c}=10^{4}$\,K (lower panels). We choose $(\alpha,\beta)=(1,1)$, a host galaxy with $R=200$ pc, ${\rm SFR}=10\,M_{\odot}$ yr$^{-1}$, $\sigma=150$ km s$^{-1}$, cloud column density after pressure equilibrium of $N_{\rm H}^{c}=10^{20}$ cm$^{-2}$ (blue lines) and $10^{21}$ cm$^{-2}$ (black lines), and the start position of the cloud to be $r_{0}=R$, $2R$, and $3R$ (solid, dashed, and dotted, respectively).  The calculation stops when $t=t_{\rm sh}$. Figure \ref{fig_Vc} shows that the cloud maximum velocities $V_{c}$ mainly depend on $T_{c}$ and $r_{0}$. Different $N_{\rm H}^{c}$ changes the cloud trajectories but not the maximum $V_{c}$. Clouds with $T_{c}=10^{3}$\,K can only be accelerated to $\Delta r/R\sim10^{-3}$ (0.2 pc) and $\sim 10^{-2}$ (2 pc) for $N_{\rm H}^{c}=10^{20}$ cm$^{-2}$ and 10$^{21}$ cm$^{-2}$, respectively. Clouds with $T_{c}\approx 10^{4}$\,K can be accelerated to $\Delta r\sim 0.1\;R$ (20 pc) for $N_{\rm H}^{c}=10^{21}$ cm$^{-2}$ and $r_{0}=3R$, a bit larger than the values of $\Delta r$ for $T_{c}=10^{3}$ K. These results are consistent with the analytic constraint given in equation (\ref{flyingdistsh}). Because of a longer survival distance, the cloud with higher $T_{c}$ can be accelerated to higher $V_{c}$. The maximum $V_{c}$ for clouds with $T_{c}=10^{3}$\,K is limited to $V_{c}\sim40$ km s$^{-1}$ or $V_{c}\sim0.04\;V_{\rm hot}$, but $V_{c}$ reaches $\sim100$\,km s$^{-1}$ or $V_{c}\sim0.1V_{\rm hot}$ for clouds with $T_{c}=10^{4}$\,K, all of which are consistent with equations (\ref{Vcsh}) and (\ref{flyingdistsh}).


Figure \ref{fig_VcContour} gives the more general result. It shows contours of maximum cloud velocity $V_{c}$ in the parameter space of $(\log\alpha,\log\beta)$. We start the calculation for clouds with $N_{\rm H}^{c}=10^{21}$\,cm$^{-2}$ at $r_{0}=R$ (left panels) and $3R$ (right panels), with $T_{c}=10^{3}$ K (upper panels) and $10^{4}$ K (lower panels), and ${\rm SFR}=10$\,M$_{\odot}$ yr$^{-1}$ with $\sigma=150$\,km s$^{-1}$. The calculations stop when $t=t_{\rm sh}$ even $t_{\rm sh}>t_{\rm evap}$. The grey regions show the parameter regime where the flow becomes radiative and the CC85 model is not valid (see \citealt{Zhang14}). We find that clouds can hardly be accelerated. The maximum value of $V_{c}$ reaches $\sim200$ km s$^{-1}$ only for $T_{c}=10^{4}$ K and $r_{0}=3R$. Otherwise $V_{c}$ is always below 100 km s$^{-1}$. Note that $V_{c}$ slightly depends on $(\alpha,\beta)$, which is different from the analytic estimate in equation (\ref{Vcsh}). This is because of gravity: for fixed $T_{c}$ and $r_{0}$, higher thermalization efficiency $\alpha$ general gives higher $V_{c}$. The critical lines of $V_{c}=0$ are given by equation (\ref{NHc_constraint1}). Overall, the entire lower left region of each panel produces no positive acceleration for the clouds because the ram pressure force does not exceed the gravitational force.

Note that because $V_{c}$ is an increasing function of $M_{\rm h}$ (equation \ref{Vcsh}), clouds with larger starting position can be accelerated to higher $V_{c}$. For example, for clouds with $r_{0}=10R$ with $R=200$ pc and $N_{\rm H}^{c}=10^{20}$ cm$^{-2}$, we find that $V_{c}$ reaches $\sim200-300$ km s$^{-1}$. This result is consistent with numerical simulations (\citealt{SB15}). Although very large values of $r_{0}$ might be reasonable for nearby halo gas or clouds over run by the hot wind after escaping the galaxy, in this paper we focus on clouds accelerated out of the host galaxy ($r_{0}\leq 3R$).




\subsection{Magnetic Fields and Large $\kappa$}\label{Sec_kappa}

In Sections \ref{Sec_Anal} and \ref{Sec_Fiducial} we assume the pressure equilibrium condition is $P_{c}=P_{\rm hot}(1+M_{\rm hot})$ (equations \ref{balance1} and \ref{balance2}), and show that the most important timescales determining the terminal velocity of cool clouds is the cloud shredding timescale (equations \ref{shtime}, \ref{constraintsh} and \ref{Vcsh}). Magnetic fields may change the structure of clouds, and potentially suppress the cloud shredding instability. We compare the thermal pressure $P_{c}$ with the magnetic pressure inside the cloud, and find that if
\begin{equation}
B\geq B_{\rm crit}=1.7\,\textrm{mG}\;P_{0}^{1/2}(1+M_{\rm h})^{1/2}\alpha^{1/4}\beta^{1/4}R_{200 \rm pc}^{-1}\textrm{SFR}_{1}^{1/2},
\end{equation}
the magnetic pressure dominates over the thermal pressure inside the cloud. Although this value of the internal cloud field is very large compare to normal star-forming galaxies and starbursts (\citealt{Thompson06}), a strong field may be generated in the rapidly cooling shock with the hot wind that initially establishes pressure equilibrium.

Magnetic fields may also suppress the cloud shredding and the KH instability, and yield a larger value of $\kappa$. Recent magnetohydrodynamic simulations show that $\kappa$ may be larger than the value of 4 implied by high-resolution hydrodynamical simulations because of cloud magnetization (e.g., \citealt{McCourt15})\footnote{\cite{McCourt15} show that a magnetic field in a hot wind may also enhance the ram pressure force by a factor of $\sim(1+V_{A}^{2}/V_{\rm hot}^{2})$, where $V_{A}$ is the Alfv\'{e}n speed in the wind. Setting $V_{\rm A}^{2}\geq V_{\rm hot}^{2}$ requires $B\geq1.2\,\textrm{mG}\,u_{0}\rho_{0}^{1/2}\alpha^{1/4}\beta^{3/4}R_{200,\rm pc}^{-1/2}\textrm{SFR}_{1}^{1/2}$
Taking $\rho_{0}\sim10^{-2}$ at $r_{0}=2R$, $u_{0}\sim1$, $\alpha\sim\beta\sim1$, this implies $B\gtrsim100\,\mu$G.}.
For this reason, although we take $\kappa=4$ in our fiducial models, the effects of larger $\kappa$ and its implications for our results should be discussed.

Equation (\ref{constraintsh}) implies that the critical value of $\kappa$ such that $t_{\rm sh}\sim t_{\rm acc}$ is
\begin{equation}
\kappa_{\rm crit}\sim390(1+M_{\rm h})^{1/6}P_{0}^{1/2}\rho_{0}^{-1/2}\alpha^{1/2} T_{c,3}^{-1/2}, \label{kappacrit}
\end{equation}
where the dimensionless factor $(1+M_{\rm h})^{1/6}P_{0}^{1/2}\rho_{0}^{-1/2}\alpha^{1/2}\sim0.61-0.29$ for $r_{0}=R$ to $3R$. This gives an analytic estimate of the required $\kappa$ for significant cloud acceleration. However, as discussed in Section \ref{Sec_PE} (equations \ref{Vcevap} and \ref{flyingdistevap}), saturated evaporation may play an important role in cloud destruction if cloud shredding is suppressed. The estimates in Section \ref{Sec_PE} imply that saturated conduction limits $V_{c}\lesssim100$ km s$^{-1}$. In fact, the presence of magnetic fields may simultaneously suppress both conduction and the cloud shredding (e.g., \citealt{Orlando08}). For these reasons, and because of the evaporation timescale is similar to the cloud shredding timescale, in the following we neglect cloud evaporation in our numerical experiments, and simply focus on the cloud shredding timescale.

Figure \ref{fig_Vckappa} shows contours of $V_{c}$ in the parameter space of $(\log\alpha,\log\beta)$ with larger $\kappa=20,60$ and $T_{c}=10^{2},10^{3}$ and $10^{4}$ K. For $\kappa=20$, a cloud with $T_{c}=10^{3}$ K and a hot wind with $\alpha\sim1,\beta\gtrsim 0.2$ can be accelerated to $V_{c}\gtrsim100$ km s$^{-1}$ (upper left), and clouds with $T_{c}=10^{4}$ K can be accelerated to $V_{c}\sim2000$ km s$^{-1}$ or even higher velocities with $\alpha\sim1,\beta\sim 0.2$. A larger value of $\kappa=60$ can accelerate clouds with $T_{c}=10^{3}$ K to the similar value of $V_{c}$ as clouds with $T_{c}=10^{4}$ K and $\kappa=20$. Also, clouds with $T_{c}=10^{2}$ K can be accelerated to $\gtrsim 100$ km s$^{-1}$ for $\kappa=60$ (lower right). Note that $\Delta r_{\rm sh}\propto \kappa^{2}$ in equation (\ref{flyingdistsh}), and thus for $r_{0}=R$ $(3R)$ with $R=200$ pc, we have $\Delta r\approx 0.65(50)$ pc $N_{\rm H,21}^{c}T_{c,3}$SFR$_{1}^{-1}$ for $\kappa=20$, and $\Delta r\approx 5.8$ pc (450 pc) $N_{\rm H,21}^{c}T_{c,3}$SFR$_{1}^{-1}$ for $\kappa=60$. In Section \ref{Sec_Cases} we compare these results with the observed cool cloud velocities.

\cite{McCourt15} showed that for magnetized clouds $\kappa$ is sufficiently large that cool clouds may become co-moving with the hot wind, and $\Delta r$ thus approaches infinity. In Section \ref{Sec_Cases} we also return to this issue.

\section{Case Studies}\label{Sec_Cases}

Here we compare the model of RPA of cool clouds by hot winds with some observations of individual starbursts, including M82, dwarf starbursts, LIRGs and ULIRGs.


\subsection{M82}\label{Sec_M82}

M82 is perhaps the most well-studied starburst galaxy in the local Universe. The total 8-1000 $\mu$m infrared luminosity of M82 $L_{\rm IR}\simeq5.6\times10^{10}L_{\odot}$ (\citealt{Sanders03}) corresponds to a SFR of $\sim5-10\,M_{\odot}$ yr$^{-1}$ (\citealt{OConnell78}; \citealt{Kennicutt98}; \citealt{FS03}; \citealt{Strickland04a}; \citealt{Elbaz07}; \citealt{SH09}; \citealt{Panuzzo10}), depending on the assumed IMF. The projected velocities of the cool or warm outflow are from $40-200$ km s$^{-1}$ in molecular emission (H$_{2}$, \citealt{Veilleux09}; SiO, \citealt{M82SiO01}; CO, \citealt{Walter02}), and $\sim100$ km s$^{-1}$ in the Na D absorption lines (\citealt{Schwartz04}), to a higher value of $\sim600$ km s$^{-1}$ for warm H$\alpha$ clumps (\citealt{Lehnert96}; \citealt{Shopbell98}). \cite{SH09} modeled the physical properties of the SN-driven hot wind based on the best currently available observations of M82.  They found that the hard X-ray observations constrain the hot wind to have $\dot{M}_{\rm hot}\sim1.4-3.6\,M_{\odot}$ yr$^{-1}$ ($\beta\sim0.1-0.6$), efficient thermalization ($\alpha\sim1$), and an implied asymptotic hot wind velocity of $V_{\rm hot}\sim1500-2000$\,km s$^{-1}$.

As expected from our analytic estimates, we find that our fiducial model ($\kappa=4$) is unable to explain the observed cool cloud velocities. In our calculations we take the total SFR of M82 as $10$\,M$_{\odot}$ yr$^{-1}$, and adopt clouds of temperature $T_{c}=10^{4}$\,K for H${\alpha}$ emission, $T_{c}=10^{3}$\,K for Na D absorbers, and $T_{c}=100$\,K for molecular emitters (see Figures \ref{fig_Vc}, \ref{fig_VcContour}). For our fiducial parameters, the maximum cloud velocities are always below $\sim100$ km s$^{-1}$ for $T_{c}=10^{3}$\,K, and below $\sim200$ km s$^{-1}$ for $T_{c}=10^{4}$\,K.

Increasing the lifetime of clouds  --- e.g., by making $\kappa$ arbitrarily large in equations (\ref{shtime}) and  (\ref{constraintsh}) as discussed in Section \ref{Sec_kappa} might solve this problem. For $\alpha=1$, $\beta=0.5$, $r_{0}=2R$ and $\kappa\sim23$, we find that clouds with $T_{c}=10^{3}$\,K are accelerated to $\sim140$ km s$^{-1}$, and clouds with $T_{c}=10^{4}$\,K reach $\sim600$ km s$^{-1}$. However, the flying distances of clouds are $\Delta r=30$\,pc and 1.1\,kpc for $T_{c}=10^{3}$\,K and $T_{c}=10^{4}$\,K, which are inconsistent with large multi-kpc extend of the emission and absorption from observations. On the other hand, if $\kappa$ is large enough that the clouds become co-moving with the hot flow, as in the magnetized cloud simulations of \cite{McCourt15}, $\Delta r$ becomes large enough to match observations. However, in this case, $V_{c}=V_{\rm hot}\sim 1500-2000$ km s$^{-1}$ and the cool cloud velocities are then too high to match observations  (e.g., \citealt{Leroy15}). We conclude that the acceleration profile, radial extent, and asymptotic velocity of cool clouds may be used as a strong constraint on any models of ram pressure acceleration.



\begin{figure}
\centerline{\includegraphics[width=10.5cm]{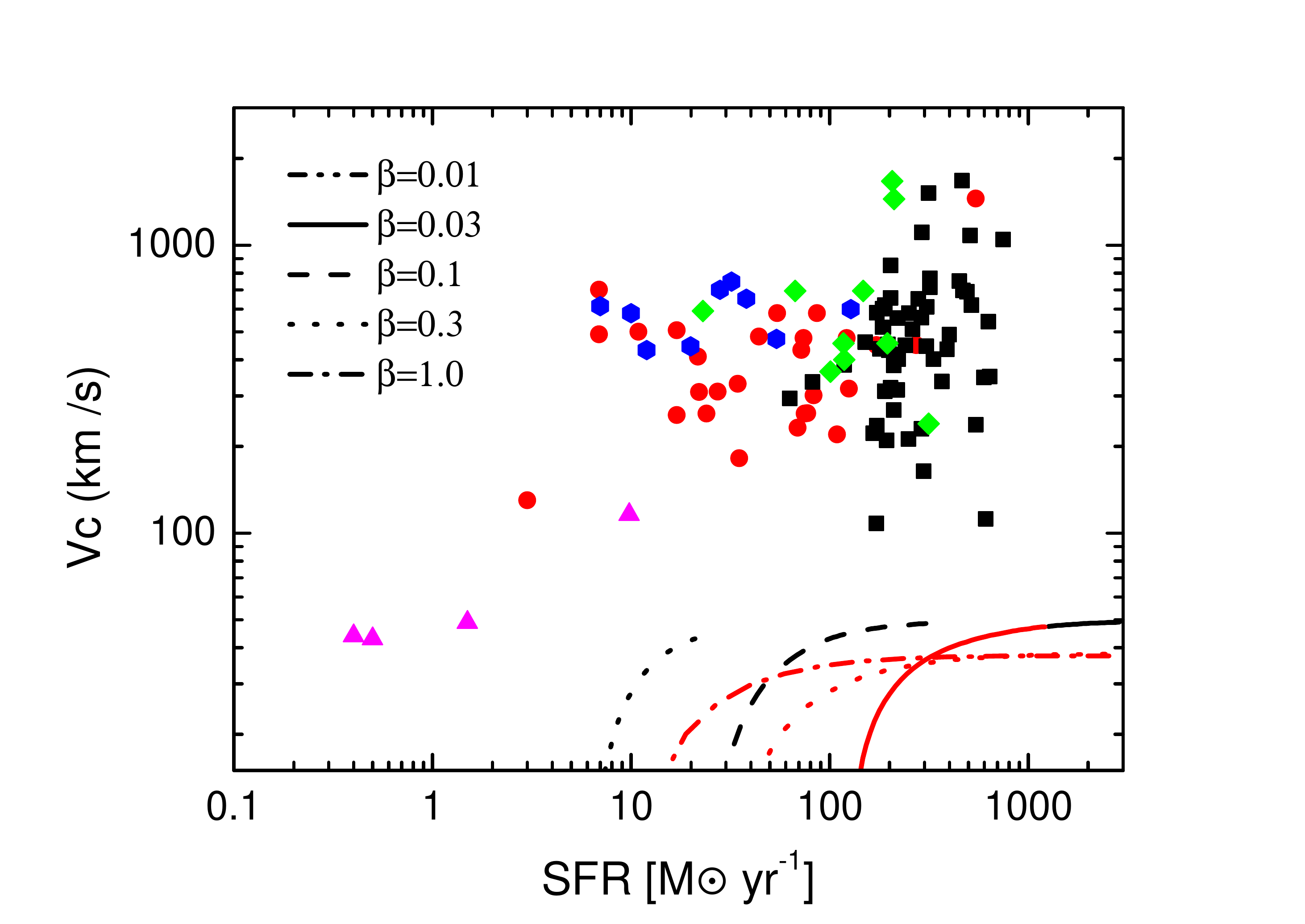}}
\caption{The relation between $V_{c}$ and SFR (combined equations \ref{rpa1}, \ref{equRc1} or \ref{equRc2}, and \ref{LxhotvsLxdiffuse}) for $\alpha$ ranging from $10^{-2}$ to 2, but with fixed $\beta=0.01$, 0.03, 0.1, 0.3 and 1, black lines are for adiabatic winds, and red lines are formally in the radiative region and the model breaks down. The parameter set is taken as $r_{0}=2R$, $R=200$ pc, $N_{\rm H}^{c}=10^{21}$ cm$^{-2}$, $T_{c}=10^{3}$ K, $f_{d}=0.1$, $\sigma=150$ km s$^{-1}$ and $\kappa=4$. The galactic outflow data are the maximum velocities taken from dwarf starbursts (\textit{triangles}, \citealt{Schwartz04}), LIRGs (\textit{circles}, \citealt{Heckman00}; \citealt{Rupke05b}), ULIRGs (\textit{squares}, \citealt{Rupke02,Rupke05b}; \citealt{Martin05}), AGN ULIRGs (\textit{diamonds}, \citealt{Rupke05c}), High-$z$ star forming (\textit{hexagons}, \citealt{Weiner09}; \citealt{Erb12}; \citealt{Kornei13}).}\label{fig_VcSFR}
\end{figure}

\subsection{Dwarf Starbursts }\label{Sec_Dwarf}

The typical outflow velocities of dwarf starburst galaxies are in the range of $V_{c}\sim20-200$\,km s$^{-1}$ (\citealt{Marlowe95}; \citealt{Martin98}; \citealt{Schwartz04}; \citealt{Keeney06}). In particular, the Na D absorbers in the sample of \cite{Schwartz04} (NGC 1569, NGC 4214, NGC 4449) have low velocities $V_{c}\sim40-50$ km s$^{-1}$, which may be explained by the CC85 model combined with the RPA scenario and additional observational constraints on the diffuse X-ray emission from these systems. We search for cool cloud wind solutions in these systems over a wide range of $\alpha$ and $\beta$, and assuming cool cloud properties as follows: $r_{0}=2R$, $T_{c}=10^{3}$\,K, $N_{\rm H}^{c}=10^{21}$\,cm$^{-2}$. We find that clouds in the three dwarf starbursts can never be accelerated to $40-50\,$km s$^{-1}$ if we use the fiducial destruction timescale in equation (\ref{shtime1}). However, we still find solutions for some dwarf starbursts if we use a slightly longer destruction timescale $\kappa=6$ instead of 4 that $t_{\rm sh}=6 t_{\rm cc}^{\rm th}\sqrt{1+M_{\rm h}}$.\footnote{The timescale is from \cite{SB15} $t_{25}=6\;t_{\rm cc}^{\rm th}\sqrt{1+M_{\rm h}}$, where $t_{25}$ means that $25\%$ of the cloud is below 2/3 of the initial cloud density.}

NGC 1569 has a diffuse X-ray luminosity $\simeq1.4\times 10^{38}$ erg s$^{-1}$ (\citealt{Ott05}) and we take ${\rm SFR}=0.4$\,M$_{\odot}$ yr$^{-1}$, $R= 100$\,pc and (very low) $\sigma=25$\,km s$^{-1}$ (\citealt{Stil02}; \citealt{Ott05}; \citealt{Pasquali11}). We find that with $\alpha\sim 1$ and $\beta\sim 1$ reproduces the observed cloud velocities and the observed X-ray luminosity, where we have calculated the band-dependent X-ray emission from the wind using the same method as in \cite{Zhang14}. The value of $\beta$ is consistent with \cite{Martin02}. Using equation (\ref{flyingdistsh}) we estimate the cloud flying distance $\Delta r\sim20 - 30$\,pc above the galaxy. We suggest that the spatial distribution and radial profile of acceleration of cool clouds could be used to further constrain the wind parameters in NGC 1569.

On the other hand, the hot wind parameters $(\alpha,\beta)$ for NGC 4449 required to yield clouds that reach $\sim40-50$\,km s$^{-1}$ produce too low X-ray emission and are inconsistent with observation. For example, the (very large) values of $\alpha\sim1.0$ and $\beta\sim1.0$ we calculate are needed to produce the cool cloud velocities, yield a hot and dense wind with an integrated X-ray luminosity of $L_{X}^{\rm 2-10\,keV}\sim7.5\times10^{37}$\,ergs s$^{-1}$, much lower than the upper limit to the diffuse X-ray emission observed ($L_{X}^{\rm 2-10\,keV}\simeq1.4\times10^{39}$\,ergs s$^{-1}$; \citealt{Bogdan11}).

Finally, due to the large gravitational potential in NGC 4214 with a value of $\sigma\sim100$ km\,s$^{-1}$ (\citealt{Thronson88}; \citealt{Schwartz04}; \citealt{DOnghia08}), clouds in NGC 4214 cannot be accelerated to the observed $V_{c}\sim40-50$ km s$^{-1}$.

In short, even though the observed cool cloud velocities are low in dwarf starbursts, we conclude that the CC85 model combined with RPA scenario can only explain some of them (e.g., NGC 1569). In addition, the spatial distribution and radial profile of acceleration of cool clouds, if observed, could be used to further constrain the wind parameters.

\subsection{LIRGs and ULIRGs}\label{Sec_ULIRGs}

Surveys of Na D absorption lines show cool gas outflows in LIRGs and UILRGs with an average velocity at the line center of $300-400$\,km s$^{-1}$, and projected maximum velocities (average velocity at center plus one-half the velocity width) up to $\sim1000$\,km s$^{-1}$.

We assume a fraction $f_{d}\lesssim 1$ of the observed total emission in X-rays from star-forming galaxies is due to a putative hot wind fluid, and then we ask whether such a flow can accelerate cool clouds to the observed velocities and physical scales. Because star-forming galaxies obey a mean linear $L_X-{\rm SFR}$ relation, we take the relation from \cite{Mineo14}
\begin{equation}
L_{X,\rm diffuse\;(0.5-8\,\rm keV)}=4.0\times10^{39}\;f_{d}\;{\rm erg\;s}^{-1}\frac{\textrm{SFR}}{(M_{\odot}\;{\rm yr}^{-1})}.\label{diffuse}
\end{equation}
Thus $f_{d}=1$ is the observed mean relation between total X-ray emission and SFR. We assume that this fraction of the total band-dependent observed emission in equation (\ref{diffuse}) is due to the hot wind (\citealt{Zhang14}):
\begin{equation}
L_{X,\rm hot}=L_{X, \rm diffuse},\label{LxhotvsLxdiffuse}
\end{equation}
where $L_{X, \rm hot}$ is the X-ray emission from the hot wind.
Equation (\ref{LxhotvsLxdiffuse}) is a function of $(\alpha,\beta)$, $R$, SFR and $f_{d}$. If we combine equation (\ref{LxhotvsLxdiffuse}) with the set of cloud acceleration equations (\ref{rpa1}), (\ref{equRc1}), and (\ref{equRc2}), the cloud velocity $V_{c}$ can be calculated as a function of the SFR for a given parameter set of $(\alpha,\beta)$, $f_{d}$, and cloud parameters $r_{0}$, $N_{\rm H}^{c}$ and $T_{c}$. The relation between $V_{c}$ and the SFR in the model can then be compared with the data from observations with the hope of constraining, ruling out, or providing evidence for the model.

We can calcuate the maximum value of $V_{c}$ (equations \ref{rpa1}, \ref{equRc1} and \ref{equRc2}) and SFR (equation \ref{LxhotvsLxdiffuse}) as functions of $\alpha$, $\beta$, $f_{d}$, and $R$, and compare the calculated $V_{c}-$SFR relation with observations. Figure \ref{fig_VcSFR} gives examples of the fiducial model. Data on the maximum outflow velocities are taken from surveys of Na D or Mg II absorption lines (\citealt{Heckman00}; \citealt{Rupke02}; \citealt{Schwartz04}; \citealt{Martin05}; \citealt{Rupke05b, Rupke05c}; \citealt{Weiner09}; \citealt{Erb12}; \citealt{Kornei13}). Since we do not know X-ray fluxes for all systems, we assume $f_{d}=0.1$. As in our previous examples, we see that hot winds in LIRGs and ULIRGs cannot accelerate cool gas to the observed velocities in the Na D surveys for our fiducial parameters. Changing other parameters including $T_{c}$, $r_{0}$, $f_{d}$ and $R$ does not change our results quantitatively. For example, assuming $f_{d}=1$ or $R=1$\,kpc, we find that $V_{c}$ is always below $\sim100$ km s$^{-1}$. However, similar to our calculations for M82 in Section \ref{Sec_M82}, larger $\kappa$ (see Section \ref{Sec_kappa}) could mitigate this conclusion.

\section{Conclusions}\label{Sec_Conclusions}

The cool gas with temperatures from $T_{c}\sim10^2$ to $10^{4}$\,K seen in emission and absorption in galactic winds may be accelerated by the ram pressure of hot winds driven by overlapping supernovae (SNe) within rapidly star-forming galaxies. We have used analytic estimates and semi-analytic models to study the acceleration and destruction of cool gas clouds as a function of both hot wind and cool cloud properties. Our main conclusions are as follows.

(1) We find that over a very broad range of parameters cool clouds always establish pressure equilibrium with the hot flow before being accelerated (equations~\ref{constraintcrushing} \& \ref{constraintexpand}; Fig.~\ref{fig_timeconstraint}).

(2) We derive a critical condition on the mass loading efficiency $\beta$ (equation~\ref{parameter2}) such that clouds in pressure equilibrium are accelerated before destruction by the cloud shredding timescale (equation~\ref{constraintsh}).  For our fiducial assumptions about the timescale $t_{\rm sh}$ (equation~\ref{shtime}), clouds with $T_c\lesssim10^{4}$\,K are destroyed before significant acceleration and these clouds do not reach velocities comparable to that of the hot wind (equations~\ref{constraintsh}, \ref{Vcsh}, \ref{flyingdistsh}).

(3) We compare the gravitational force ($F_{\rm grav}$) with the ram pressure force ($F_{\rm ram}$), deriving an Eddington-like limit for $F_{\rm ram}\geq F_{\rm grav}$ as a function of cloud and host galaxy properties (equations~\ref{NHc_constraint1}, \ref{constraintgrav0}, \ref{NHconstraint1}; Fig.~\ref{fig_NHconstraint}). If we take an initial cloud to be compressed by the ram pressure of the hot wind and come into pressure equilibrium with the hot wind, we show that the initial column density of launched clouds must be less than $\sim10^{21}$\,cm$^{-2}$ for outward acceleration with $T_{c}=10^{3}$ K and $R=200$ pc (equations~\ref{NHconstraint1}, \ref{NHconstraint2} and Fig.~\ref{fig_NHconstraint}). Higher $T_{c}$ or $R$ can increase the upper bound to $N_{\rm H}\lesssim10^{22}$ cm$^{-2}$. These estimates depend sensitively on the properties of clouds.

(4) The timescale for cloud shredding $t_{\rm sh}$ plays the most important role in determining the final velocities of clouds $V_{c}$. For $T_{c}\sim10^{3}$\,K, as might be appropriate for absorption studies of the Na D lines which have been widely observed in surveys of galactic outflows, $V_{c}$ is limited to $\lesssim100$\,km s$^{-1}$ by cloud shredding (equation~\ref{Vcsh}; Figs.~\ref{fig_Vc} and \ref{fig_VcContour}), and the clouds are accelerated and destroyed very near their starting positions $r_{0}$ (equation~\ref{flyingdistsh}), potentially in conflict with observations. Similarly, warm clouds ($T_{c}\sim10^{4}$ K) and molecular clouds ($T_{c}\lesssim100$ K) cannot be accelerated by hot flows to observed velocities over virtually any range in parameter space. However, as we show in Section \ref{Sec_kappa} (Fig. \ref{fig_Vckappa}), $V_{c}$ can be significantly higher if the magnetic cloud shredding timescale ($t_{\rm sh}$) is increased by a factor of $\sim 15$ and $\sim 5$ for $T_{c}=10^{3}$ K and $10^{4}$ K respectively due to cloud magnetization (\citealt{McCourt15}), as long as conductive evaporation can be neglected. We derive a critical $\kappa_{\rm crit}$ such that $t_{\rm sh}\sim t_{\rm acc}$ (equation \ref{kappacrit}) as a guide for current and future simulations.

We then compare our models with observations of outflows in M82, dwarf starbursts, LIRGs and ULIRGs. We combine the X-ray luminosities of star-forming galaxies with the scenario of ram pressure acceleration (RPA) of cool clouds by assuming a diffuse hot wind X-ray luminosity that contributes a fraction $f_{d}$ to the total X-ray luminosity of star-forming galaxies (equation \ref{diffuse}). As expected from our analytic investigation, this picture fails to produce velocities high enough to match observations, expect for some dwarf starbursts (e.g., NGC 1569). However, as in our previous examples, the cool clouds may well be explained if the cloud shredding time is much longer than implied by hydrodynamical simulations. Note, though, that even in cases where $\kappa$ is 15 times larger and clouds reach large $V_{c}$, the spatial extend and the acceleration profile may be inconsistent with observations. The later thus provides a particularly powerful probe of the wind acceleration mechanism. 

Overall we conclude that individual cool clouds with $T_{c}\lesssim10^{4}$\,K accelerated by ram pressure of a hot wind are not likely to match observed cool gas outflows. However, other cloud acceleration and formation scenarios or wind driving mechanisms may explain the observed properties of cool gas outflows in rapidly star-formation galaxies. In our model, we assume individual dense clouds with an initial scale of $\Delta R\sim N_{\rm H}^{\rm i}/2n_{\rm H}^{\rm i}\simeq0.2\,$pc $N_{\rm H,21}^{\rm i}/n_{\rm H,3}^{\rm i}$ and a mass of $M_{c}\simeq 0.4\,M_{\odot}\,N_{\rm H,21}^{\rm i,3}n_{\rm H,3}^{\rm i,-2}$ to be accelerated in the hot wind. In reality, it may be that giant cool gas shells with masses of $\sim10^{8}-10^{10}M_{\odot}$ on kpc scales are pushed out by the ram pressure of the hot wind, and that these shells eventually fragment, littering the hot outflow with cool gas clouds that are then accelerated on larger scales and mix with the hot wind. On the other hand, radiation pressure-driven winds may also be able to accelerate cool clouds (\citealt{MQT05}; \citealt{KT12,KT13}; \citealt{Hopkins12}; \citealt{Zhang14}; \citealt{Thompson15}). For example, in \cite{Zhang14} we showed that radiation pressure driving is one possibility to explain the SFR$-V_{c}$ relation observed in Na D surveys. Another possibility is that outflows are driven by the pressure of cosmic rays (e.g., \citealt{Everett08}; \citealt{Socrates08}; \citealt{Jubelgas08}; \citealt{Booth13}; \citealt{SB15}). 

In summary, entrainment and ram pressure acceleration by a hot wind are strongly constrained. Clouds in only a narrow range of initial column densities can be accelerated, and are shredded rapidly at small distances from their launch radii and at relatively low velocities. This calls into question the prevailing picture where the gas probed by absorption and emission is thought to be entrained and ram pressure accelerated by the hot wind. Cool clouds can be accelerated to the observed velocities only if magnetic fields in the clouds are sufficiently important to prolong the lifetime of the clouds and suppress the evaporation, but even in this case the spatial extend and acceleration profile should be tested against observations of resolved systems like M82  (Section \ref{Sec_M82}).

\section*{Acknowledgments}
We thank the referee Evan Scannapieco for his very useful comments that have allowed us to improve our paper. D.Z. thanks Crystal Martin, Sylvain Veilleux, Evan Scannapieco and Claude-Andr\'{e} Faucher-Gigu\`{e}re, and T.A.T. thanks Tim Heckman for a number of stimulating discussions. E.Q. thanks Ryan O'Leary and Mike McCourt for useful conversations. This work is supported in part by NASA grant \# NNX10AD01G. T.A.T is supported in part by NSF \#1516967. E.Q. is supported in part by NASA ATP Grant 12-ATP12-0183, a Simons Investigator award from the Simons Foundation, the David and Lucile Packard Foundation, and the Thomas Alison Schneider Chair in Physics. N.M. is supported in part by NSERC of Canada and by the Canada Research Chair program.

\end{document}